\global\def\draftcontrol{0}

   \def\versionno{ LG orientifolds }

\catcode`\@=11

\expandafter\ifx\csname draftcontrol\endcsname\relax\global\def\draftcontrol{0} 
\fi 

{\count255=\time\divide\count255 by 60 
\xdef\hourmin{\number\count255} 
\multiply\count255 by-60\advance\count255 by\time 
\xdef\hourmin{\hourmin:\ifnum\count255<10 0\fi\the\count255}} 
\def\draftdate{\number\month/\number\day/\number\year\ \ \ \hourmin } 

\newcommand\makepapertitle{\par

  \begingroup 
    \renewcommand\thefootnote{\@fnsymbol\c@footnote}%
    \def\@makefnmark{\rlap{\@textsuperscript{\normalfont\@thefnmark}}}%
    \long\def\@makefntext##1{\parindent 1em\noindent 
            \hb@xt@1.8em{%
                \hss\@textsuperscript{\normalfont\@thefnmark}}##1}%
     \newpage 
     \global\@topnum\z@   
     \@makepapertitle 
     \thispagestyle{empty}\@thanks 
  \endgroup 
  \setcounter{footnote}{0}%
  \global\let\thanks\relax 
  \global\let\makepapertitle\relax 
  \global\let\@makepapertitle\relax 
  \global\let\@thanks\@empty 
  \global\let\@author\@empty 
  \global\let\@date\@empty 
  \global\let\@title\@empty 
  \global\let\title\relax 
  \global\let\author\relax 
  \global\let\date\relax 
  \global\let\and\relax 
  \def\version{\let\version\@version\@gobble} 
} 
\def\@makepapertitle{%
  \newpage 
   \ifnum\draftcontrol=1 {} 
   \version\versionno 
   \vskip 5em%
   \else 
   \hfill\hbox to 3cm {\parbox{4cm}{\@pubnum}\hss}%
   \vskip 5em%
   \fi 
   \begin{center}%
   \let \footnote \thanks 
      {\hskip -0\textwidth \hbox to 1\textwidth%
        {\centerline{\Large\bf{\noindent\@title}}}}%
     \vskip 2em%
     {\normalsize
       \lineskip .5em%
       \begin{tabular}[t]{c}%
         \@author 
       \end{tabular}\par}%
     \vskip 1em%
     {\@bstract}%
     \end{center}%
     \vfill
     \@date%
     \vskip 1.5em%
   \par 
} 

\gdef\@pubnum{} 
\def\pubnum#1{%
  \gdef\@pubnum{#1}} 

\gdef\@bstract{} 
\def\Abstract#1{%
  \gdef\@bstract{%
   \parbox{\textwidth-0pc}{%
   \centerline{\bf Abstract}\penalty1000 
   \noindent
   \renewcommand\baselinestretch{1.0} 
   {#1}}} 
} 

\gdef\@email{}
\def\email#1{%
   \gdef\@email{%
   Email: {\tt #1}}
}

\def\ps@paper{\let\@mkboth\@gobbletwo%
     \ifnum\draftcontrol=1 
        \def\@oddfoot{\hbox to \textwidth{\tiny \versionno \hfil\tiny\draftdate}%
        \hskip -\textwidth \hbox to \textwidth{\hfil\rm\thepage\hfil}}%
     \else\def\@oddfoot{\hbox to \textwidth{\hfil\rm\thepage\hfil}} 
     \fi 
     \let\@evenfoot\@oddfoot 
} 

\def\body{\clearpage 
          \pagestyle{paper} 
        } 
\newenvironment{acknowledgments}{%
\vskip 3.25ex 
\noindent {\bf Acknowledgments} 
} 

\def\@version#1{\ifnum\draftcontrol=1 
\typeout{}\typeout{#1}\typeout{} 
\vskip3mm\centerline{\hbox{\fbox{\normalsize{\tt DRAFT -- #1 -- } 
                   {\draftdate}}}}\vskip3mm 
\fi} 
\let\version\@version 
\long\def\eqlabel#1{\ifnum\draftcontrol=1 
                    \tag@false  
                    \tag*{(\theequation) \hbox to -0.2cm{\hspace{0cm}\small{#1}\hss}} 
                    \refstepcounter{equation}  
                    \edef\@currentlabel{\theequation} 
                    \ltx@label{#1}          
                    \else 
                    \label{#1} 
                    \fi 
                    } 
\let\st@bibitem\@bibitem 
\let\st@lbibitem\@lbibitem 
\ifnum\draftcontrol=1 
  \def\@bibitem#1{%
    \st@bibitem{#1}\a@@label{#1}\ignorespaces} 
  \def\@lbibitem[#1]#2{%
    \st@lbibitem[#1]{#2}\a@@label{#2}\ignorespaces} 
  \def\a@@label#1{%
    \gdef\a@lab{\smash{\normalfont\small#1}} 
    \ifvmode 
      \if@inlabel 
        \global\setbox\@labels\hbox{%
          \llap{\a@lab\let\a@lab\relax 
                \kern\@totalleftmargin\kern\marginparsep}%
          \box\@labels}%
      \fi 
    \fi} 
\fi 

\documentclass[12pt,letterpaper]{article} 

\usepackage{amsmath,bm,amsfonts,amssymb,array,calc,amsthm,rotating}
\usepackage{epsfig,psfrag} 
\usepackage[nosort]{cite} 
\usepackage{graphicx}
\usepackage{color}
\usepackage[colorlinks=false]{hyperref}

\renewcommand\baselinestretch{1.25} 
\renewcommand\section{\@startsection {section}{1}{\z@}%
                                   {-3.5ex \@plus -1ex \@minus -.2ex}%
                                   {2.3ex \@plus.2ex}%
                                   {\normalfont\large\bfseries}} 
\renewcommand\subsection{\@startsection{subsection}{2}{\z@}%
                                   {-3.25ex\@plus -1ex \@minus -.2ex}%
                                   {1.5ex \@plus .2ex}%
                                   {\normalfont\normalsize\bfseries}} 
\renewcommand\subsubsection{\@startsection{subsubsection}{3}{\z@}%
                                   {-3.25ex\@plus -1ex \@minus -.2ex}%
                                   {1.5ex \@plus .2ex}%
                                   {\normalfont\normalsize\it}} 
\renewcommand\paragraph{\@startsection{paragraph}{4}{\z@}%
                                   {-3.25ex\@plus -1ex \@minus -.2ex}%
                                   {1.5ex \@plus .2ex}%
                                   {\normalfont\normalsize\bf}} 
\renewcommand\subparagraph{\@startsection{subparagraph}{5}{\z@}%
                                   {-1.25ex\@plus -1ex \@minus -.2ex}%
                                   {0ex \@plus .2ex}%
                                   {\normalfont\normalsize\it}} 

\numberwithin{equation}{section}

\long\def\@makecaption#1#2{%
  \vskip\abovecaptionskip
  \sbox\@tempboxa{{\bf #1:} #2}%
  \ifdim \wd\@tempboxa >\hsize
    {\small\bf #1:} {\small #2}\par
  \else
    \global \@minipagefalse
    \hb@xt@\hsize{\hfil\box\@tempboxa\hfil}%
  \fi
  \vskip\belowcaptionskip}

\renewcommand*\l@section[2]{%
  \ifnum \c@tocdepth >\z@
    \addpenalty\@secpenalty
    \addvspace{.5em \@plus\p@}%
    \setlength\@tempdima{1.5em}%
    \begingroup
      \parindent \z@ \rightskip \@pnumwidth
      \parfillskip -\@pnumwidth
      \leavevmode \bfseries
      \advance\leftskip\@tempdima
      \hskip -\leftskip
      #1\nobreak\hfil \nobreak\hb@xt@\@pnumwidth{\hss #2}\par
    \endgroup
  \fi}
\renewcommand*\l@subsection{\addvspace{.0em \@plus\p@}\@dottedtocline{2}{1.5em}{2.3em}}
\renewcommand*\l@subsubsection{\addvspace{-.2em \@plus\p@}\@dottedtocline{3}{3.8em}{3.2em}}

\def\hepth#1{\href{http://xxx.arxiv.org/abs/hep-th/#1}{{arXiv:hep-th/#1}}}

\def\mathag#1{\href{http://xxx.arxiv.org/abs/math.AG/#1}{{arXiv:math.ag/#1}}}
\def\alggeom#1{\href{http://xxx.arxiv.org/abs/alg-geom/#1}{{arXiv:alg-geom/#1}}}

\definecolor{refcol}{rgb}{0.2,0.2,0.8}
\definecolor{eqcol}{rgb}{.6,0,0}
\definecolor{purple}{cmyk}{0,1,0,0}

\gdef\@citecolor{refcol}
\gdef\@linkcolor{eqcol}
\def\colorlinkspurple{\gdef\@urlcolor{purple}}
\def\colorlinksblue{\gdef\@urlcolor{blue}}
\def\colorlinksred{\gdef\@urlcolor{red}}

\def\ie{{\it i.e.}} 
\def\eg{{\it e.g.}}

\def\revise#1       {\raisebox{-0em}{\rule{3pt}{1em}}%
                     \marginpar{\raisebox{.5em}{\vrule width3pt\ 
                     \vrule width0pt height 0pt depth0.5em 
                     \hbox to 0cm{\hspace{0cm}{%
                     \parbox[t]{4em}{\raggedright\footnotesize{#1}}}\hss}}}}

\def\calh         {{\cal H}} 
 
\def\calj         {{\cal J}}

\def\caln         {{\cal N}} 
 
\def\calp         {{\cal P}} 
 
\def\calr         {{\cal R}}

\def\complex      {{\mathbb C}} 
 
\def\projective   {{\mathbb P}} 
\def\rationals    {{\mathbb Q}} 
\def\reals        {{\mathbb R}} 
\def\zet          {{\mathbb Z}} 

\def\del          {\partial} 
\def\delbar       {\bar\partial} 
\def\ee           {{\it e}} 
\def\ii           {{\it i}} 
 
\def\tr           {{\rm Tr}}

\def\id           {{\rm id}}

\newcommand\topa[2]{\genfrac{}{}{0pt}{2}{\scriptstyle #1}{\scriptstyle #2}}

\def\sqr#1#2{{\vcenter{\vbox{\hrule height.#2pt   
 \hbox{\vrule width.#2pt height#1pt \kern#1pt 
 \vrule width.#2pt}\hrule height.#2pt}}}}

\DeclareFontFamily{U}{rsf}{}
\DeclareFontShape{U}{rsf}{m}{n}{
  <5> <6> rsfs5 <7> <8> <9> rsfs7 <10-> rsfs10}{}
\DeclareMathAlphabet\Scr{U}{rsf}{m}{n}

\def\str{{\mathop{\rm Str}}}
\def\Pf{\mathop{\rm Pf}}

\def\HH{{\mathfrak H}}

\def\MF{{\rm MF}}          

\def\Hom{{\rm Hom}}

\def\oHom{\hat{\rm H}{\rm om}}
\def\oEnd{\hat{\rm E}{\rm nd}}

\newcommand{\MFD}{\Scr{M\!F}}
\newcommand{\Pa}{\Scr{P}}
\newcommand{\aPa}{\Scr{A}\!\Pa}
\newcommand{\Anti}{\Scr{A}}

\newcommand{\dd}{{\rm d}}

\def\res{{\rm Res}}

\def\oGamma{\widehat\Gamma}

\newcommand{\opsi}{\overline{\psi}}

\newcommand{\bepsilon}{\overline{\epsilon}}

\newcommand{\beq}{\begin{equation}}
\newcommand{\eeq}{\end{equation}}
\newcommand{\beqa}{\begin{eqnarray}}
\newcommand{\eeqa}{\end{eqnarray}}

\newcommand{\C}{{\mathbb C}}

\catcode`\@=12 

\begin{document} 


\title{%
\parbox{\textwidth}{
\centerline{D-brane Categories for Orientifolds}
\centerline{--- The Landau-Ginzburg Case ---}}}

\pubnum{%
hep-th/0606179}
\date{June 2006}

\author{
Kentaro Hori$^{a}$ and Johannes Walcher$^{b}$ \\[0.4cm]
\it $^a$ Department of Physics, University of Toronto,\\
\it Toronto, Ontario, Canada \\[0.2cm]
\it $^b$ School of Natural Sciences, Institute for Advanced Study,\\
\it Princeton, New Jersey, USA}

\Abstract{
We construct and classify categories of D-branes in orientifolds based 
on Landau-Ginzburg models and their orbifolds. Consistency of the 
worldsheet parity action on the matrix factorizations plays the key 
role. This provides all the requisite data for an orientifold 
construction after embedding in string theory. One of our main results 
is a computation of topological field theory correlators on unoriented 
worldsheets, generalizing the formulas of Vafa and Kapustin-Li 
for oriented worldsheets, as well as the extension of these results
to orbifolds. We also find a doubling of Kn\"orrer periodicity in 
the orientifold context.
}


\makepapertitle

\body

\version\versionno

\vskip 1em

\tableofcontents
\newpage

\parskip 6pt

\section{Introduction}

Despite their importance for model building, orientifolds\cite{BiSa,Horava,GiPo}
have not been receiving the attention they deserve.
One of the reasons is the lack of a well-developed mathematical 
structure into which orientifolds can be framed. This is in sharp 
contrast with the case of D-branes for oriented strings
\cite{DLP,Polchinski},
where the structure is extensively studied both from physics and 
from mathematics.
One of the key discovery in the latter context is that the language of
category fits very well and machinery of
homological algebra can be applied effectively \cite{Kontsevich,FOOO,douglas}.
It is natural to ask whether this continues to be so also for unoriented
strings.\footnote{
It is worthwhile mentioning that in the categorification 
program of rational CFT \cite{frs}, unoriented worldsheets fit in very 
naturally, and are included from the beginning.}

In this paper, we approach this question
by simply constructing the categories that are
relevant for physics of unoriented strings.
We do this in a class of very simple and tractable backgrounds 
--- the Landau-Ginzburg models with ${\mathcal N}=2$
worldsheet supersymmetry. The realization that
matrix factorization of the superpotential
are the correct description of B-type D-branes
 has provided a simple model 
of the topological D-brane category that is at the same time concrete
 and tractable \cite{kali1,bhls,kali2,calin,hl,everybody}. 
We will not here attempt a systematic foundation of the subject
that can be applied to more general backgrounds, although
we believe that such a treatment can be rather straightforwardly given 
based on our present results.
Also, we note that Landau-Ginzburg models
have an advantage in transition from the topological realm
to the physical world, at non-geometric points in the
Calabi-Yau moduli space.

A basic program to achieve the same goal for
geometric backgrounds was presented in the first part of
\cite{talk} based on \cite{GH}.
Without orientifold, where the complex K\"ahler moduli 
analytically connect Landau-Ginzburg orbifolds and
 Calabi-Yau sigma models, the category of matrix factorizations is equivalent 
to the derived category of coherent sheaves on the underlying algebraic 
variety. This was conjectured in \cite{johannes} and the
 equivalences were constructed mathematically in \cite{orlov}
and are physically understood in \cite{HHP}.
It is very interesting to see the relation in the
orientifold models where the K\"ahler moduli are
projected to ``real'' locus.

We will also develop a technology to compute
correlation functions of topological field theory
for unoriented worldsheets, generalizing the 
formulas for oriented worldsheets by Vafa for the closed strings 
\cite{toplg} and of Kapustin-Li for open strings \cite{kali2}.
One of our main results in this paper is the construction of the 
crosscap states in Landau-Ginzburg models as well as in Landau-Ginzburg 
orbifolds where the parity is involutive only up to the
orbifold group.

The rest of the paper is organized as follows. Section \ref{general} reviews 
the worldsheet origins of B-type orientifolds of Landau-Ginzburg models 
\cite{BH2,BHHW,GH}. In particular, we explain why the action of parity
on matrix factorization is given by a ``graded transpose'', as one
could have naively anticipated: If a matrix squares to a scalar multiple
of the identity, the transposed matrix will also do this. In section
\ref{functors}, we formulate a parity as an anti-involutive functor
from the category of D-branes to itself, and then specialize to the
category of matrix factorizations in LG models and their orbifolds.
In our general classification of parities, we find the possibility 
for a ``twist by quantum symmetry'' in the definition of parity 
acting in Landau-Ginzburg orbifolds. In section \ref{category}, we 
define an orientifold category as the fixed category under the parity
functor with the additional requirement that the action of parity on
the morphism spaces be involutive. This produces several (standard) 
results about possible gauge groups on D-branes in orientifolds.
We emphasize that it {\it does not} make sense to
restrict the morphism spaces themselves to their invariant subspaces,
because this would not be compatible with the algebraic structure.
We also include a discussion of the compatibility of orientifolding
with R-charge grading, and hence D-brane stability, in the homogeneous 
case. In section \ref{state}, we compute topological crosscap correlators, 
both in theories with involutive parity (in other words, the parent theory 
is an ordinary Landau-Ginzburg, not an orbifold), as well as in orientifolds
of Landau-Ginzburg orbifolds. The derivation includes a formula for the 
parity twisted Witten index in the open string sector between a brane and 
its orientifold image. We conclude with some comments in section \ref{final}.

Some of the material in this paper was presented in \cite{talk} and 
\cite{jotalk}. We were informed by Emanuel Diaconescu of
upcoming work by the Rutgers group which treats the problem of 
orientifold of derived category in the geometric regime.

\section{Parity Actions on D-branes and Open Strings}
\label{general}

For a general discussion of parity symmetries in $\caln=2$ supersymmetric
worldsheet theories, we refer to \cite{BH2}. Here, we will first recall
a few basic facts about parities in $\caln=2$ Landau-Ginzburg models,
emphasizing the structure of the orientifold group. We will then analyze
the action of parities on D-branes and open strings, first for
 general background and next in Landau-Ginzburg models.
We find that ``graded transpose'' plays a relevant role.
At the end of the section,
we record basic notion and convention of $\zet_2$ graded linear algebra.

\subsection{Parity symmetries for closed strings}

As explained in \cite{BH2}, a B-type orientifold is defined from a parity
\begin{equation}
P = \tau \Omega
\eqlabel{P}
\end{equation}
where $\Omega$ is (B-type) (super)worldsheet parity, while $\tau$ is 
an action on target space variables. In a two-dimensional $\caln=(2,2)$
Landau-Ginzburg model, requiring that $P$ be a symmetry means in particular 
that $\tau$ should act holomorphically on chiral field variables $x$ such 
that the superpotential transforms with a minus sign
\begin{equation}
\tau^* W(x) = W(\tau x) = - W(x)
\eqlabel{WP}
\end{equation}
This minus sign is required for the bosonic Lagrangian
\begin{equation}
\int d\theta^+d\theta^- W (x)
\end{equation}
to be invariant under B-type parity dressed with $\tau$ \cite{BH2}.
The requirement that $P$ be an involutive parity can be relaxed 
if we allow the bulk theory to be a Landau-Ginzburg orbifold.
As explained by Douglas and Moore \cite{domo}, the orientifold group 
is generically an extension
\begin{equation}
\Gamma \longrightarrow \oGamma \overset{\pi}{\longrightarrow} \zet_2
\eqlabel{origroup}
\end{equation}
In other words, a parity is any element $P\in \oGamma$ such that
$\pi(P)$ is the non-trivial element of $\zet_2$, while a general
element $g$ of the orbifold group $\Gamma$ has trivial image in
$\zet_2$. To be able to gauge $\oGamma$, we need $P^2\in\Gamma$.

In such a context, the action of $P$ on the theory can have a dependence
on the twisted sector on which it is acting, as explained for example
in \cite{BHHW}. Concretely, this data is an element $\chi$ of the character
group of $\Gamma$, and when acting on a state in the sector twisted
by $g$, we include a phase factor $\chi(g)$. (This can be viewed as
discrete-torsion like phases in the definition of the Klein bottle.)

To classify such orientifolds, let us denote by $G$ the symmetry group
of $W$ generated by phase symmetries and permutations of the variables.
Although $\tau$ need not be a symmetry, the difference of two dressings 
is always in $G$. For simplicity, we will generally assume 
that $\Gamma\subset G$ is an abelian orbifold group consisting just of phase 
symmetries. Possible quantum symmetries are elements of $\Gamma^*$ and  
for abelian $\Gamma$, form a group isomorphic to $\Gamma$. The group of 
symmetries of $W/\Gamma$ is therefore
\begin{equation}
G^{\Gamma} = (G/\Gamma) \times \Gamma^* \cong G
\end{equation}
To find the possible inequivalent orientifold dressings $\tau$, 
for fixed orbifold group $\Gamma$, we must impose $P\Gamma P^{-1}
\subset \Gamma$, $P^2\in\Gamma$, as well as identify dressings
which differ by conjugation by an element $g\in G^{\Gamma}$.
Since parity reverses the orientation of a string, it maps
a string twisted by $g\in\Gamma$ to a string twisted by $g^{-1}$.
Therefore
\begin{equation}
P\chi P^{-1} = \chi^{-1} 
\eqlabel{transform}
\end{equation}
where $\chi^{-1}=\bar\chi$ is the character inverse to 
$\chi\in\Gamma^*$. As a consequence, we find that choices of
inequivalent dressings correspond to involutive elements of
$G$ commuting with $\Gamma$ as well as elements of $\Gamma^*$
$\bmod (\Gamma^*)^2$. For examples, see \cite{BHHW}.

To make this section complete, we recall that when $W$ is homogeneous,
\begin{equation}
W(\ee^{\ii\lambda q_i} x_i) = \ee^{2\ii\lambda} W(x_i)\,,
\eqlabel{homo}
\end{equation}
for some $q_i\in\rationals$, there is a standard choice of orbifold group 
$\Gamma\cong\zet_H$ (where $H$ is the smallest integer such that 
$H q_i\in2\zet$ for all $i$). $\zet_H$ is generated by $g$ corresponding 
to $\lambda=\pi$ in \eqref{homo}. There is then also a standard 
choice of parity, corresponding to $\lambda =\pi/2$ in \eqref{homo}.
We then have $\tau^2=g$, so $\tau$ actually generates the full
orientifold group. But it should be remembered that this is not
true in general: There can be elements of $\Gamma$ which are not
the square of any parity, and several parities can square to
the same element of $\Gamma$.

\subsection{Parity action on open strings --- a general story}

The purpose of this section is to show that the parity action on open
strings with
$\zet_2$ or $\zet$ graded Chan-Paton spaces
is given by {\it graded transpose} (described below).
This is extracted from the work \cite{GH}. 

In general, an oriented string boundary carries a discrete degree
of freedom represented in some complex vector space $V$,
called a Chan-Paton (CP) space. If the orientation is flipped,
the Chan-Paton space is replaced by its dual $V^*=\Hom(V,\complex)$.
Let us consider the open string worldsheet $\Sigma=\reals\times [0,\pi]$ 
where $\reals$ and $[0,\pi]$ are spanned by the time and the space 
coordinates, $t$ and $\sigma$, respectively.
We take the convention that the right boundary ($\sigma=\pi$)
is oriented in the $t$-increasing direction while the left boundary
($\sigma=0$) is oriented oppositely.
\begin{figure}[htb]
\centerline{\includegraphics{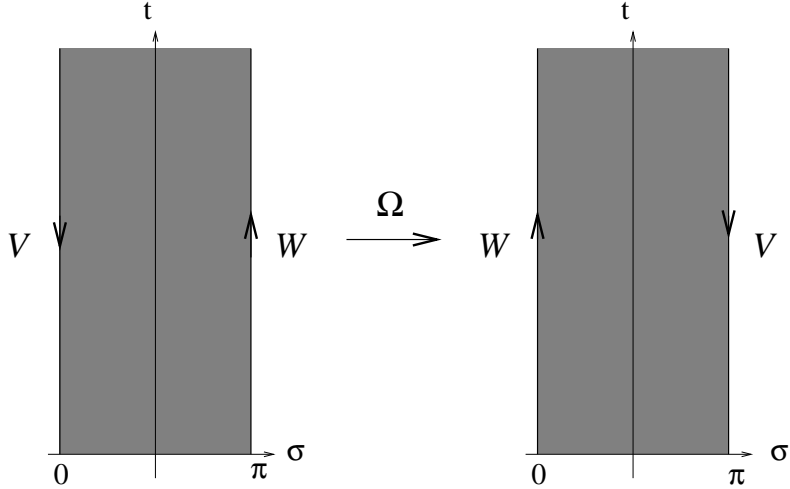}}
\caption{Orientation reversal of the open string
worldsheet}
\label{rev}
\end{figure}
Suppose that the left and the right boundary carry Chan-Paton spaces
$V$ and $W$ respectively. (See the left part of 
Figure~\ref{rev}.)
If we quantize the open string
in such a way that the increase in $t$
corresponds to positive time evolution,
the space of states include a factor (Chan-Paton factor)
\begin{equation}
{\mathcal H}^{\rm CP}=\Hom(V,W).
\end{equation}
Now, if we consider the orientation reversal,
$$
\Omega:(t,\sigma)\longmapsto (t,\pi-\sigma),
$$
the left and the right boundaries are swapped and they are
oppositely oriented compared to the standard convention we have chosen.
Then the Chan-Paton factor is
\begin{equation}
\Omega({\mathcal H}^{\rm CP})=\Hom(W^*,V^*).
\end{equation}
A parity operator includes a complex linear isomorphism
\begin{equation}
P:\Hom(V,W)\longrightarrow
\Hom(W^*,V^*).
\end{equation}
Up to automorphisms of
$V$ and $W$, a natural candidate is
the transpose, $P(\phi)=\phi^t$, defined by
$$
\langle\phi^tf,v\rangle=\langle f, \phi(v)\rangle,\qquad
\mbox{for $f\in W^*$ and $v\in V$}.
$$
However, if $V$ and $W$ are graded vector spaces, it appears more
natural to use the graded transpose $P(\phi)=\phi^T$
defined by
$$
\langle\phi^Tf,v\rangle=(-1)^{|f||\phi|}\langle f, \phi(v)\rangle.
$$
Here $|f|=0$ when $f$ is even and $|f|=1$  if $f$ is odd (similarly for
$|\phi|$). Namely, a sign appears whenever two ``fermionic''
objects swap their positions.
In what follows, we confirm this guess using the case where the
boundary degrees of freedom are complex (or even number of real)
fermions.

Suppose that the left and the right boundaries
 carry real fermions
$\xi_j$ ($j=1,...,2m$) and
$\eta_a$ ($a=1,...,2n$) respectively.
The kinetic term of these variables is
\begin{eqnarray}
S_{boundary}&=&\int_{-\infty}^{\infty}
\dd t \sum_{a=1}^{2n}{\frac i2}\eta_a{\frac{\dd}{\dd t}}\eta_a
+\int_{-\infty}^{\infty}
\dd t' \sum_{j=1}^{2m}{\frac i2}\xi_j{\frac{\dd}{\dd t'}}\xi_j
\nonumber\\
&=&\int_{-\infty}^{\infty}
\dd t \left[\sum_{a=1}^{2n}{\frac i2}\eta_a{\frac{\dd}{\dd t}}\eta_a
-\sum_{j=1}^{2m}{\frac i2}\xi_j{\frac{\dd}{\dd t}}\xi_j\right].
\label{actionS}
\end{eqnarray}
In the first line, $t'$ is the time coordinate
running in the opposite direction of $t$ ($t'=-t$).
The Chan-Paton factor on the right ({\it resp.} left) boundary is
a graded irreducible representation $M_{\eta}$ ({\it resp.} $M_{\xi}$)
of the Clifford algebra $\{\eta_a,\eta_b\}=\delta_{a,b}$
({\it resp.} $\{\xi_i,\xi_j\}=\delta_{i,j}$) which is unique up to
isomorphism.
If we quantize the open string in the usual way, where $t$ is the time,
the Chan-Paton factor 
$\Hom(M_{\xi},M_{\eta})$
must be a graded irreducible representation of
 the following anti-commutation relations:
\begin{equation}
\{\hat{\eta}_a,\hat{\eta}_b\}=\delta_{a,b},\quad
\{\hat{\xi}_i,\hat{\xi}_j\}=-\delta_{i,j},\quad
\{\hat{\eta}_a,\hat{\xi}_j\}=0.
\label{Calg}
\end{equation}
It is indeed represented on $\Hom(M_{\xi},M_{\eta})$ by
\begin{equation}
\hat{\eta}_a\phi:=\eta_a\circ \phi,
\qquad
\hat{\xi}_i\phi:=(-1)^{\phi}\phi\circ \xi_i.
\label{defac1}
\end{equation}
Note the unusual sign of the $\hat{\xi}$ anticommutators in (\ref{Calg}): 
it comes from the unusual sign of the $\xi$-kinetic term
in (\ref{actionS}). The sign factor $(-1)^{\phi}$ in
(\ref{defac1}) is required to produce 
this sign as well as to satisfy the relation
$\{\hat{\eta}_a,\hat{\xi}_j\}=0$.

Let us operate the worldsheet orientation reversal
$\Omega$ that swaps the two boundary lines.
The Chan-Paton factor is then $\Hom(M_{\eta}^*,M_{\xi}^*)$
since now $\eta_a$ is on the left boundary 
and $\xi_i$ on the right, both with the unconventional orientation.
The kinetic term remains the same
as (\ref{actionS}) and they must still obey the anticommutaion relations
(\ref{Calg}), which must be represented
in $\Hom(M_{\eta}^*,M_{\xi}^*)$.
The unique choice (up to isomorphism) is
\begin{equation}
\hat{\eta}_a\phi^*:=(-1)^{\phi^*}\phi^* \circ \eta_a^T,
\qquad
\hat{\xi}_i\phi^*:=\xi_i^T\circ \phi^*.
\label{defac2}
\end{equation}
Here $(-)^T$ is the graded transpose, which is needed to obtain
the correct sign for
the anticommutation relation.
Since we simply move $\eta$ from right to left and
$\xi$ from left to right,
the parity operator
$$
P:\Hom(M_{\xi},M_{\eta})\longrightarrow \Hom(M_{\eta}^*,M_{\xi}^*)
$$
must obey
$$
P^{-1}\hat{\eta}_aP=\hat{\eta}_a,\quad
P^{-1}\hat{\xi}_iP=\hat{\xi}_i.
$$
Namely, $P$ must commute with $\hat{\eta}$ and $\hat{\xi}$.
The graded transpose
\begin{equation}
P(\phi)=\phi^T
\end{equation}
satisfies this condition, and it is unique 
up to scalar multiplication.

\subsection{The case of Landau-Ginzburg models}
\label{thecase}
\newcommand{\bX}{\overline{X}}
\newcommand{\bx}{\overline{x}}

We now consider the parity operation for open strings
in Landau-Ginzburg model.
A B-type D-brane in the Landau-Ginzburg model
is specified by a matrix factorization
of the superpotential $W(x^1,...,x^n)$,
$$
Q(x)=\left(\begin{array}{cc}
0&f(x)\\
g(x)&0
\end{array}\right),
\quad Q(x)^2=W(x){\bf 1}_{2r}.
$$
It enters into the super-Wilson line factor 
$P\exp\left(-i\int_{\partial\Sigma}{\mathcal A}_t\dd t\right)$
for the tachyon configuration
${\bf T}=Q+Q^{\dag}$ of the space-filling brane-antibrane system
\cite{kali1}.
To be explicit, it is
\cite{blin,TTU,KraLar}
$$
{\mathcal A}_t={\frac 1 2}\{Q(x),Q(x)^{\dag}\}
+{\frac 12}\sum_{i=1}^n\psi_i{\frac{\partial}{\partial x_i}}Q(x)
+{\frac 12}\sum_{i=1}^n{\frac{\partial}{\partial\bx_i}}Q(x)^{\dag}
\opsi_i
$$
where $\psi_i$ are the
superpartners of the boundary values of $x_i$.
Provided $Q(x)^2=W(x){\bf 1}$ is satisfied,
its supersymmetry variation 
is given by
\begin{equation}
\delta{\mathcal A}_t\,=\,
-{\rm Re}\left(\,\sum_{i=1}^n
\bepsilon \psi_i{\frac{\partial}{\partial x_i}} W\,\right)
-i{\mathcal D}_t\Bigl(\bepsilon Q+\epsilon Q^{\dag}\Bigr)
+i\Bigl(\dot{\bepsilon}Q+\dot{\epsilon}Q^{\dag}\Bigr),
\end{equation}
where ${\mathcal D}_tX=\dot{X}+i[{\mathcal A}_t,X]$.
The first term cancels the supersymmetry variation of the bulk action
(the Warner term).
The second term is a total derivative
when inserted in the path-ordered exponentials.
The last term shows that $Q$ and $Q^{\dag}$
provides the boundary contribution to the supercharges.

One important point is that $Q$ and $Q^{\dag}$ are secretly fermionic ---
a sign must appear when a fermionic field
 $\psi_i, \opsi_i$ passes through them. We must keep track of
such signs when we find out the parity action on matrix factorizations.
Although it is possible to do so, 
we take another route: we consider matrix factorizations
that can be realized using boundary fermions where
the sign and the statistics are completely under control.
For this purpose, it is enough to consider one-by-one factorizations
where the super-Wilson-line factor is produced
by the path-integral over a single {\it complex} boundary fermion
$\eta,\overline{\eta}$ 
with the action (see e.g. \cite{bRG})
\begin{equation}
S_{\it right}=\int_{-\infty}^{\infty}\dd t\left[
i\overline{\eta}{\frac{\dd}{\dd t}}\eta
-{\frac 12}|g(x)|^2-{\frac 12}|f(x)|^2
-{\rm Re}\Bigl(\overline{\eta}\psi^i\partial_ig(x)
+\eta \psi^i\partial_if(x)\Bigr)
\right]_{\sigma=\pi},
\label{act2}
\end{equation}
and supersymmetry variation
\begin{equation}
\delta\eta=-\epsilon \overline{f(x)}-\overline{\epsilon}g(x),\qquad
\delta\overline{\eta}=-\overline{\epsilon}f(x)-\epsilon\overline{g(x)}.
\end{equation}
The subscript ``right'' emphasizes that it applies to
the brane at the right boundary of the string which is oriented
in the same direction as the time $t$.
On the left boundary, which is oriented oppositely to $t$,
the action and the variation are given by
\begin{equation}
S_{\it left}=\int_{-\infty}^{\infty}\dd t\left[
-i\overline{\xi}{\frac{\dd}{\dd t}}\xi
-{\frac 12}|\tilde{g}(x)|^2-{\frac12}|\tilde{f}(x)|^2
+{\rm Re}\Bigl(\overline{\xi}\psi^i\partial_i\tilde{g}(x)
+\xi \psi^i\partial_i\tilde{f}(x)\Bigr)
\right]_{\sigma=0},
\label{act3}
\end{equation}
\begin{equation}
\delta\xi=\epsilon \overline{\tilde{f}(x)}-\overline{\epsilon}\tilde{g}(x),
\qquad
\delta\overline{\xi}=-\overline{\epsilon}\tilde{f}(x)
+\epsilon\overline{\tilde{g}(x)},
\label{var3}
\end{equation}
where $\xi, \overline{\xi}$ are a complex fermion, and
$(\tilde{f},\tilde{g})$ is another factorization of $W$.
\footnote{$\xi$ and $\overline{\xi}$ obey the non-standard reality
relation
$(\xi)^{\dag}=-\overline{\xi}$ because it has the ``wrong'' sign kinetic term
(it we had quantized by taking $-t$ as the time, it would have had
the standard relation). Thus, the two variations in (\ref{var3})
are consistent with reality, and also the ``real part'' in
(\ref{act3}) must be with respect to such a reality.}
With these two boundary terms, the boundary part of the B-type
supersymmetry charge is given by
\begin{equation}
Q_B=\Bigl[f(x)\eta+g(x)\overline{\eta}\Bigr]_{\sigma=\pi}
-\Bigl[\tilde{f}(x)\xi+\tilde{g}(x)\overline{\xi}\Bigr]_{\sigma=0}.
\end{equation}
In particular, using (\ref{defac1}), we find that
this acts on the CP factor $\Hom(M_{\xi},M_{\eta})$
by
$$
Q_B\phi=Q\circ \phi-(-1)^{\phi}\phi\circ \tilde{Q}.
$$
Now, let us perform the parity $\tau\Omega$. Then the left and the right
boundary lines are swapped, and $(f,g)$ and $(\tilde{f},\tilde{g})$
are replaced by $(\tau^*f,\tau^*g)$ and $(\tau^*\tilde{f},\tau^*\tilde{g})$.
As a consequence, the boundary part of the B-type supercharge is now
\begin{equation}
Q_B=\Bigl[f(\tau x)\eta+g(\tau x)\overline{\eta}\Bigr]_{\sigma=0}
-\Bigl[\tilde{f}(\tau x)\xi+\tilde{g}(\tau x)\overline{\xi}\Bigr]_{\sigma=\pi}
\end{equation}
Using (\ref{defac2}), we see that it acts on the Chan-Paton factor 
$\Hom(M^*_{\eta},M^*_{\xi})$ as
\begin{equation}
Q_B\phi^*=-\tau^*\tilde{Q}^T\circ \phi^*+(-1)^{\phi^*}\phi^*\circ
\tau^*Q^T,
\end{equation}
where $(-)^T$ is the graded transpose. We see that
the matrix factorization on the right and the left boundaries are
now $-\tau^*\tilde{Q}^T$ and $-\tau^*Q^T$.

Let us summarize what we found:
Under the simple orientation reversal of the worldsheet,
a matrix factorization transforms as
\begin{equation}
Q(x)\longrightarrow -Q(\tau x)^T,
\label{parQ}
\end{equation}
while open string wavefunctions transform as
\begin{equation}
\phi(x)\longmapsto \phi(\tau x)^T.
\label{parst}
\end{equation}
In these expressions, $(-)^T$ stands for the graded transpose.

\subsection{Dual and transpose for graded vector spaces}
\label{conventions}

Since the graded transpose plays an important role, we record
below a few basic notions and conventions regarding $\zet_2$ graded
linear algebra. In LG models, it is more convenient
to regard the Chan-Paton spaces as (free) modules over the ring 
$\complex[x_1,...,x_n]$, rather than complex vector spaces.
But we describe everything in the category of complex vector spaces
because generalization to modules over the ring is straightforward.

So Let $M$ be a $\zet_2$ graded (finite-dimensional, complex) vector space
with grading operator $\sigma$ which acts as $1$ on even elements 
and $-1$ on odd elements. The dual vector space $M^*=\Hom(M,\complex)$ 
is naturally graded by $\langle\sigma^*f,e\rangle=\langle f,\sigma 
e\rangle$. Here, we use $\langle\cdot,\cdot\rangle$ to denote the
natural pairing $M^*\times M\to \complex$. It is worthwhile 
emphasizing that despite appearances, this pairing is neither 
symmetric nor anti-symmetric in any sense. In fact, there is no
preference between thinking of the pairing as a map $M^*\times M\to 
\complex$ or as a map $M\times M^*\to \complex$.

Let us now consider a linear map $A:M_1\to M_2$ between graded vector 
spaces $(M_1,\sigma_1)$ and $(M_2,\sigma_2)$.
If $A$ is even (resp. odd), we denote $|A|=0$ (resp. $|A|=1$) modulo 2.
We define the ({\bf graded}) {\bf transpose} of $A$ as a linear map 
$A^T:M_2^*\to M_1^*$ by setting
\begin{equation}
\langle A^T f_2,e_1\rangle=(-1)^{|A||f_2|}\langle f_2,A e_1\rangle,
\eqlabel{gradedtrans}
\end{equation}
if $A$ is even or odd.
If $A$ is a linear isomorphism $A^T$ is also an isomorphism.
The transpose of the inverse and the inverse of the transpose are
related by
$$
(A^{-1})^T=(-1)^{|A|}(A^T)^{-1}.
$$
If $A$ is even, they are the same and we sometimes
use the shorthand notation 
$A^{-T}=(A^{-1})^T=(A^T)^{-1}$.

We wish to emphasize that the definition \eqref{gradedtrans} in fact
constitutes a choice, and we could have equally well defined
\begin{equation}
\langle {}^T\!\!A f_2,e_1\rangle = (-1)^{|A||e_1|}\langle f_2,A e_1\rangle
\eqlabel{alternate}
\end{equation}
The latter choice is in fact the natural one if we prefer to pair a 
vector space and its dual in the opposite order. Of course, ${}^T\!\!A= 
(-1)^{|A|}A^T$, and one can check that none of our results depend on
the choice. Having said that, we will fix the graded transpose defined by 
\eqref{gradedtrans}. Let us describe some of its properties.

\noindent
\underline{\bf Matrix representation}~
Let us choose basis of $M_1$ and $M_2$ such that even elements come 
first and odd elements follow them, and suppose that $A$ is expressed 
with respect to this basis as a matrix
\begin{equation}
A=\left(\begin{array}{cc}
a&b\\
c&d\\
\end{array}\right)
\eqlabel{matrix}
\end{equation}
where $a$ maps even to even, $b$ maps odd to even, etc.
Then, the matrix representation of the graded transpose of $A$
with respect to the dual basis is
\begin{equation}
A^T=\left(\begin{array}{cc}
a^T&-c^T\\
b^T&d^T
\end{array}\right),
\eqlabel{grtrans}
\end{equation}
where $a^T$, $b^T$, $c^T$, $d^T$ are the transpose matrices of
$a$, $b$, $c$, $d$.

\noindent
\underline{\bf Change of grading}~
Graded transpose of course
depends on the gradings $\sigma_1$ and $\sigma_2$.
Let us denote by $A^{T_{\sigma_1,\sigma_2}}$
when we want to show the grading. Then, for change of grading,
the graded transpose changes as follows
\begin{equation}
A^{T_{\sigma_1,-\sigma_2}}=-\sigma_1^TA^{T_{\sigma_1,\sigma_2}},
\qquad
A^{T_{-\sigma_1,\sigma_2}}=A^{T_{\sigma_1,\sigma_2}}\sigma_2^T,
\quad
A^{T_{-\sigma_1,-\sigma_2}}
=\sigma_1^TA^{T_{\sigma_1,\sigma_2}}\sigma_2^T.
\eqlabel{changeg}
\end{equation}
Note that $\sigma$ is always even and thus its transpose
$\sigma^T$ does not depend on the grading used.

\noindent
\underline{\bf Composition}~
Graded transpose obeys an interesting property under composition.
Let $B:M_1\to M_2$ and $A:M_2\to M_3$ be linear maps.
Suppose each of them is even or odd, so that $|A|$ and $|B|$
makes sense.
Then we have
\begin{equation}
(AB)^T=(-1)^{|A||B|}B^TA^T.
\eqlabel{compose}
\end{equation}
Since this is important, let us record the proof here:
\begin{equation}
\begin{split}
\langle (AB)^Tf_3,e_1\rangle_1
&=(-1)^{|AB||f_3|}\langle f_3,ABe_1\rangle_3
\\
&=(-1)^{|AB||f_3|+|A||f_3|}\langle A^Tf_3,Be_1\rangle_2
\\
&=(-1)^{|AB||f_3|+|A||f_3|+|B||A^Tf_3|}
\langle B^TA^Tf_3,e_1\rangle_1.
\end{split}
\end{equation}
Using $|AB|=|A|+|B|$, $|A^Tf_3|=|A^T|+|f_3|=|A|+|f_3|$,
we find that the sign that appears on the right hand side
is $(-1)^{|B||A|}$. This shows \eqref{compose}.

\noindent
\underline{\bf Tensor product}~
It is also important to study and fix our convention on
tensor products.
Let $(M_1,\sigma_1)$ 
and $(M_2,\sigma_2)$
be graded vector spaces.
The tensor product $M_1\otimes M_2$ has a natural grading
$\sigma_{12}=\sigma_1\otimes \sigma_2$.
Its dual is identified with $M_1^*\otimes M_2^*$
whose dual is in turn identified with $M_1\otimes M_2$, so that
\begin{equation}
\langle a_1 \otimes a_2,b_1\otimes b_2\rangle_{12}
=(-1)^{|a_2||b_1|}\langle a_1,b_1\rangle_1\langle a_2,b_2\rangle_2
\eqlabel{deftensor}
\end{equation}
holds
for $a_1,b_1\in M_1\oplus M_1^*$ and 
$a_2,b_2\in M_2\oplus M_2^*$.
For linear maps $A_1:M_1\to N_1$ and $A_2:M_2\to N_2$
of graded vector spaces, their tensor product
map is defined by
$$
(A_1\otimes A_2)(a_1\otimes a_2)=(-1)^{|A_2||a_1|}
(A_1a_1)\otimes (A_2a_2).
$$
One can show that
the transpose map $N_1^*\otimes N_2^*\to M_1^*\otimes M_2^*$ of
this is given by
\begin{equation}
(A_1\otimes A_2)^T=A_1^T\otimes A_2^T.
\end{equation}

\noindent
\underline{\bf Double dual, double transpose}~
For a graded vector space $M$, the dual of $M^*$ 
is isomorphic to $M$. An isomorphism $\iota:M\to M^{**}$
(which we call {\it the canonical isomorphism}) is defined by 
the property
\begin{equation}
\langle \iota(v),f\rangle
=(-1)^{|v||f|}\langle f,v\rangle.
\label{iota}
\end{equation}
It of course depend on the grading $\sigma$, and we sometimes write
$\iota_{M,\sigma}$ or $\iota_{\sigma}$. Definition \eqref{iota} comes
with a few oddities (compared to, let's say, $\langle\tilde\iota(v),
f\rangle=\langle f,v\rangle$). If we choose a basis of $M$ where even 
elements come first, with respect to it and its dual-dual basis, $\iota$ 
is represented as
\begin{equation}
\iota=\left(\begin{array}{cc}
1&0\\
0&-1
\end{array}\right).
\eqlabel{dualdual}
\end{equation}
If $A:M_1\to M_2$ is a linear map of graded vector spaces, 
the map $A^{TT}:M_1^{**}\to M_2^{**}$ is related to $A$ by
\begin{equation}
A^{TT}=\iota_2A\iota_1^{-1}.
\label{ATT}
\end{equation}
If we flip the grading, the canonical isomorphism
changes by sign
\begin{equation}
\iota_{-\sigma}=-\iota_{\sigma}.
\end{equation}

\section{Parities as Functors}
\label{functors}

As is well-known \cite{Kontsevich,FOOO}, in a theory of oriented strings
the collection of all D-branes
forms a category ${\mathcal C}$ (to be precise,
in a slightly generalized sense).
Objects of ${\mathcal C}$ are D-branes, $B_i$, the space of
morphisms between two objects, $\Hom(B_1,B_2)$,
is the space of states of the open string
stretched from the brane $B_1$ to the brane $B_2$.
The composition of morphisms dictates the process of two strings
joining into one.

In this language, a parity operation can be regarded as an
{\it anti-involution} of that category.
By definition, it is a
contravariant functor
\begin{equation}
{\mathcal P}:{\mathcal C}\to {\mathcal C},
\end{equation}
whose square is isomorphic to the identity,
\begin{equation}
\calp \circ \calp\cong {\id}_{\mathcal C}.
\end{equation}
Namely, to each brane $B$ a brane ${\mathcal P}(B)$ is assigned
as its parity image.
For each pair of branes $(B_1,B_2)$ there is a
linear isomorphism
$$
{\mathcal P}:\phi\in\Hom(B_1,B_2)\longmapsto
{\mathcal P}(\phi)\in\Hom({\mathcal P}(B_2),{\mathcal P}(B_1)),
$$
mapping states of the open string from $B_1$ to $B_2$ to
states of the open string from ${\mathcal P}(B_2)$ to ${\mathcal P}(B_1)$,
in a way compatible with the joining process.
Furthermore, if the parity is operated twice on a brane
the result is isomorphic
to the original brane,
${\mathcal P}({\mathcal P}(B))\cong B$,
in a way consistent with everything.

In what follows, we define such parity functors in the categories
of D-branes associated with a LG model.
We start with describing the categories themselves.
(This is a review.)

\subsection{The category of matrix factorizations}

As discussed above, the data to specify a B-brane in the LG model
is a matrix factorization $Q(x)$
of the superpotential $W(x)$. This can be interpreted as an open string
tachyon configuration on a stack of equal numbers of space filling 
branes and anti-branes. We regard this as a triple
\begin{equation}
(M,\sigma,Q)
\end{equation}
where $M$ is a free module over the polynomial ring
$S=\C[x_1,...,x_n]$, with $\zet_2$ grading
$\sigma$, and $Q$ is an odd endomorphism of $M$ which squares to
$W$ times the identity of $M$.
The (truncated off-shell) space of open string states
 between two branes determined by 
matrix factorizations $(M_1,\sigma_1,Q_1)$ and $(M_2,\sigma_2,Q_2)$ 
is the space of homomorphisms of the modules $\Hom_S(M_1,M_2)$. The 
supercharge action is represented by $\dd \phi=Q_2\phi-\sigma_2\phi
\sigma_1 Q_1$. Matrix factorizations of $W$ with $\bigl(\Hom_S(M_1,M_2),
\dd\bigr)$ as morphism spaces form a differential graded category which 
we denote by $\MFD(W)$. 
Its homotopy category $\MF(W)$, obtained by simply restricting morphisms to be
$\dd$-cohomology classes, is triangulated
\cite{buchweitz,bonkap,orlov1}.

In the case of LG orbifolds, the data for 
a brane includes an even representation $\rho$ 
of $\Gamma$ on $M$, satisfying the condition
\begin{equation}
\rho(g) Q(g x) \rho(g)^{-1} = Q(x) \qquad \text{for $g\in\Gamma$}
\eqlabel{orbifold}
\end{equation}
This is of course nothing else but the original orbifold construction
of \cite{domo}, actually with the simplification that all branes,
being space filling, are invariant under $\Gamma$. Therefore, we 
only have an action on the Chan-Paton space $M$ together with the action on 
closed string variables.
Thus, a B-brane in the LG orbifold is specified by the data
\begin{equation}
(M,\sigma,Q,\rho).
\end{equation}
For a pair branes given by such data, $(M_1,\sigma_1,Q_1,\rho_1)$
and $(M_2,\sigma_2,Q_2,\rho_2)$, the space of open strings
between them is given by the $\Gamma$-invariant
homomorphisms $\Hom_S(M_1,M_2)^{\Gamma}$ with respect to the
action $g\in\Gamma:\phi\mapsto \rho_2(g)g^*\phi\rho_1(g)^{-1}$.
These data form a differential graded category $\MFD_{\Gamma}(W)$
and we denote its homotopy category by $\MF_{\Gamma}(W)$ which is
again triangulated. The group $\Gamma^*$ of quantum symmetries 
acts as the auto-equivalence of these categories by
$(M,\sigma,Q,\rho)\mapsto (M,\sigma,Q,g^*(\rho))$, where
$(g^*(\rho))(g)=  g^*(g)\,\rho(g)$ for $g^*\in\Gamma^*$,
$g\in\Gamma$.

\subsection{The parity functors}

Let us now define the parity functors on these categories.
We first consider the case without orbifold.
An example is already found in section~\ref{thecase}: a matrix 
factorization $Q(x)$ is sent to $-Q(\tau x)^T$ and a morphism
$\phi(x)$ is mapped to $\phi(\tau)^T$, 
see Eqn~(\ref{parQ})
and Eqn.~(\ref{parst}).
Thus, we propose to define a functor $\Pa:\MFD(W)\to\MFD(W)$ by
\begin{equation}
\begin{split}
(M,\sigma,Q)&\longmapsto(M^*,\sigma^T,-\tau^*Q^T),\\
 \Hom_S(M_1,M_2)\ni\phi & \longmapsto
\tau^*\phi^T\in \Hom_S(M_2^*,M_1^*).
\end{split}
\eqlabel{Pa1}
\end{equation}
It meets the conditions:\\
$\bullet$ $\Pa(M,\sigma,Q)$ is a matrix factorization of $W$
\begin{equation}
\begin{split}
(-\tau^*Q^T)^2&=
\tau^*Q^T\tau^*Q^T=-(\tau^*Q\tau^*Q)^T
=-(\tau^*W\cdot{\rm id}_M)^T\\
&=-(-W)\cdot{\rm id}_{M^*}=W\cdot{\rm id}_{M^*},
\end{split}
\end{equation}
where we have used \eqref{compose} in the second equality
and also \eqref{WP} in the fourth.\\
$\bullet$ It commutes with the supercharge:
For $\phi\in \Hom_S(M_1,M_2)$ which is either even or odd,
$\sigma_2\phi\sigma_1=(-1)^{|\phi|}\phi$, we have
\begin{equation}
\begin{split}
\dd \Scr{P}(\phi)&=(-\tau^*Q_1^T)\tau^*\phi^T
-(-1)^{|\phi|}\tau^*\phi^T(-\tau^*Q_2^T) \\
&=-(-1)^{|\phi|}(\tau^*\phi\tau^*Q_1)^T
+(\tau^*Q_2\tau^*\phi)^T\\
&=\tau^*(Q_2\phi-(-1)^{|\phi|}\phi Q_1)^T=\Pa(\dd\phi)
\end{split}
\label{commu}
\end{equation}
$\bullet$
It is compatible with the composition: For $\phi\in \Hom_S(M_1,M_2)$
and $\psi\in \Hom_S(M_2,M_3)$, their composition
$\psi\phi\in\Hom_S(M_1,M_3)$ is mapped to  
\begin{equation}
\Pa(\psi\phi)=\tau^*(\psi\phi)^T=
(-1)^{|\psi||\phi|}\tau^*\phi^T\tau^*\psi^T
=(-1)^{|\psi||\phi|}\Pa(\phi)\Pa(\psi).
\end{equation}
$\bullet$ The square of $\Pa$ does
\begin{equation}
\begin{split}
\Pa^2&:(M,\sigma,Q)\longmapsto (M^{**},\sigma^{TT},\tau^*(\tau^*Q^T)^T)
=(M^{**},\sigma^{TT},Q^{TT}),\\
\Pa^2&:\Hom_S(M_1,M_2)\ni\phi\longmapsto 
\phi^{TT}\in \Hom_S(M_1^{**},M_2^{**}).
\end{split}
\end{equation}
Using the relation \eqref{ATT}, we see that the canonical isomorphism 
$\iota:M\to M^{**}$ provides an isomorphism of $\Pa^2$ to
the identity,
$$
\iota:\Pa\stackrel{\cong}{\longrightarrow}{\rm id}_{\MFD(W)}.
$$
Thus, $\Pa$ is an anti-involution of the differential graded category
$\MFD(W)$. As a consequence, it descends to a functor of the homotopy
category $\MF(W)$. Namely, $\Pa$ defines a map of $\dd$-cohomology classes
by the property (\ref{commu}).

One may obtain other parity functors by composing the above $\Pa$ 
with some other auto-equivalence of $\MFD(W)$. For example, $\MFD(W)$ 
has {\it antibrane functor} $\Anti$ that sends $(M,\sigma,Q)$ to
$(M,-\sigma,Q)$. Let us examine the composition $\aPa:=\Anti\circ\Pa$ 
$$
\aPa:(M,\sigma,Q)\longmapsto (M^*,-\sigma^T,-\tau^*Q^T).
$$
The transpose ``$T$'' is with respect to the grading $\sigma$.
It is easy to see that it is a matrix factorization of $W$, and that
$\dd\aPa(\phi)=\aPa(\dd\phi)$ for any open string
state $\phi$. The square of $\aPa$ does
\begin{equation}
\begin{split}
(\aPa)^2&:(M,\sigma,Q)\longmapsto (M^{**},\sigma^{TT},-Q^{TT}),\\
(\aPa)^2&: \Hom_S(M_1,M_2)\ni\phi\longmapsto (\sigma_2\phi\sigma_1)^{TT}
\in \Hom_S(M_1^{**},M_2^{**})
\end{split}
\end{equation}
where we have used \eqref{changeg}. We see that $\iota\circ\sigma:M\to M^{**}$ 
provides an isomorphism of $(\aPa)^2$ to the identity.

\subsection{LG orbifolds}

Let us next consider the orbifold of the LG model by a finite abelian 
group $\Gamma$ that preserves the superpotential, $g^*W=W$ for all 
$g\in\Gamma$. We suppose that an abelian extension of $\zet_2$ by $\Gamma$
$$
1\longrightarrow
 \Gamma \longrightarrow \oGamma \longrightarrow \zet_2\longrightarrow 1,
$$
also acts on the variables, so that elements $\tau\in \oGamma$ outside 
of $\Gamma$ flip the sign of $W$, $\tau^*W=-W$. One can consider a 
parity symmetry associated with the group $\oGamma/\Gamma$.

We would like to find parity actions on B-branes and open string states.
The requirement is as before --- we seek anti-involutions
of the categories
$\MFD_{\Gamma}(W)$ and
$\MF_{\Gamma}(W)$.
Let us choose any odd element $\tau\in\oGamma\setminus\Gamma$, and
consider
\begin{equation}
\Pa(\tau):(M,\sigma,Q,\rho)\longmapsto(M^*,\sigma^T,-\tau^*Q^T,\rho^{-T}).
\end{equation}
It is easy to see that $\phi\mapsto\tau^*\phi^T$ maps $\Gamma$-invariants
to $\Gamma$-invariants;
\begin{equation}
\Pa(\tau):\phi\in\Hom_S(M_1,M_2)^{\Gamma}\longmapsto
\tau^*\phi^T\in \Hom_S(M_2^*,M_1^*)^{\Gamma}.
\end{equation}
The square of $\Pa(\tau)$ is isomorphic to the identity by the canonical 
isomorphism $\iota:M\to M^{**}$. 
It commutes with the supercharge and thus the functor descends to the homotopy 
category.
For any $g\in \Gamma$, $\Pa(\tau g)$
is isomorphic to $\Pa(\tau)$ by $\rho(g):M\to M$.
However, it may appear unpleasant that we need to make a choice
of $\tau\in\oGamma\setminus\Gamma$.
This worry actually disappears when we appropriately
define the D-brane category in the orientifold. See Section~\ref{oo}.

One may dress the action on $\rho$ by a character
$\chi:\Gamma\longrightarrow \complex^{\times}$,
\begin{equation}
\Pa_{\chi}(\tau):(M,\sigma,Q,\rho)
\longmapsto(M^*,\sigma^T,-\tau^*Q^T,\chi\rho^{-T}).
\label{Pachi}
\end{equation}
We also have $\aPa_{\chi}(\tau)=\Anti\circ\Pa_{\chi}(\tau)$.

\section{Category of D-branes in Orientifolds}
\label{category}

So far, we have discussed parity as an anti-involution of
the category of D-branes. In constructing an orientifold
background in string theory, we intend to promote parity from
a global symmetry to a gauge symmetry. In particular, we want to 
include into the background a configuration of D-branes
that is invariant by the parity, and classify open string states
with specific transformation properties
under the parity.

This motivates us to consider a new category of D-branes, whose objects are
invariant brane configurations or simply invariant branes.
Let ${\mathcal C}$ be a category of D-branes with a parity functor
${\mathcal P}$. 
An invariant brane is a brane $B$ with an isomorphism
\begin{equation}
U:{\mathcal P}(B)\longrightarrow B.
\end{equation}
For a pair of invariant branes, $(B_1,U_1)$ and $(B_2,U_2)$,
there is a linear map
\begin{equation}
P:\Hom(B_1,B_2)\stackrel{{\mathcal P}}{\longrightarrow}
\Hom({\mathcal P}(B_2),{\mathcal P}(B_1))
\stackrel{U_1\cdot (?)\cdot U_2^{-1}}{\longrightarrow}
\Hom(B_2,B_1).
\label{statetra}
\end{equation}
We require that this linear map be an involution
\begin{equation}
P\circ P= {\id}_{\Hom(B_1,B_2)}.
\label{involution}
\end{equation}
In order to impose orientifold projection, we want $P^2$ to be {\it 
literally} the identity operator of the space of states 
$\Hom(B_1,B_2)$. (It is not enough for $\calp^2$ to be isomorphic
to the identify.) This is a rather strong condition
on the collection of invariant objects $(B,U)$.
It is possible that, for each functor ${\mathcal P}$,
there are several categories consisting of 
invariant branes that are mutually
compatible in the sense that (\ref{involution}) holds for any pair.
As for the morphisms in the new categories,
the best choice is to keep all morphisms from before the 
orientifold, as we will discuss below.

The purpose of this section is to show that one can indeed define such
categories and to classify them, in Landau-Ginzburg models
and their orbifolds.

\subsection{The categories 
\texorpdfstring{$\MFD_{{\Pa}}^{\,\epsilon}(W)$
and $\MF_{{\Pa}}^{\,\epsilon}(W)$}{}}

We start with the unorbifolded Landau-Ginzburg model
 with an involution $x\mapsto \tau x$ such that $W(\tau x)=-W(x)$.
As the parity functor, let us first 
take ${\mathcal P}=\Pa$ defined in (\ref{Pa1}).
An invariant brane is a quadruple $(M,\sigma,Q,U)$ where
$(M,\sigma,Q)$ is a matrix factorization  of $W(x)$ and
$U$ is a linear isomorphism
\begin{equation}
U:M^*\longrightarrow M,
\eqlabel{Ui}
\end{equation}
such that
\begin{equation}
\begin{split}
&U\sigma^TU^{-1}=\sigma,\\[0.2cm]
&U(-\tau^*Q^T)U^{-1}=Q.
\end{split}
\eqlabel{condi}
\end{equation}
For a pair of such branes,
$(M_1,\sigma_1,Q_1,U_1)$ and $(M_2,\sigma_2,Q_2,U_2)$,
the parity transformation of the open string states
is defined according to (\ref{statetra}):
\begin{equation}
P:\phi\in\Hom_S(M_1,M_2)\longmapsto
U_1\tau^*\phi^TU_2^{-1}\in \Hom_S(M_2,M_1).
\label{Pdef}
\end{equation}
The square of $P$ is given by
\begin{eqnarray}
P^2(\phi)&=&
U_2\tau^*(U_1\tau^*\phi^TU_2^{-1})^TU_1^{-1}
=U_2U_2^{-T}\phi^{TT}U_1^TU_1^{-1}
\nonumber\\
&=&U_2U_2^{-T}\iota_2\phi \iota_1^{-1}U_1^TU_1^{-1},
\nonumber
\end{eqnarray}
where $\phi^{TT}=\iota_2\phi\iota_1^{-1}$ is used in the last step
(see (\ref{ATT})).
We require that the right hand side equals $\phi$ itself
for any $\phi\in \Hom_S(M_1,M_2)$. By Schur's Lemma,
this means that
$U_2U_2^{-T}\iota_2$ and $U_1U_1^{-T}\iota_1$ are the same 
matrix which is proportional to the identity.
Thus, we find that, in a category of mutually compatible branes,
we have
\begin{equation}
UU^{-T}\iota=\epsilon\cdot {\rm id}_M
\label{defep}
\end{equation}
where $\epsilon$ is a constant that is independent of the brane in
that category. Using $U=\epsilon \iota^{-1}U^T$ twice and
employing the relation $\iota^{-1}U^{TT}\iota^{-T}=U$,
we find that
$$
\epsilon^2=1.
$$
Thus, we obtain a category parametrized by a sign $\epsilon$, which we denote
by $\MFD_{\!\Pa}^{\,\epsilon}(W)$.
As for the morphisms, we decide to keep everything. Namely
the space of morphisms from
$(M_1,\sigma_1,Q_1,U_1)$ to $(M_2,\sigma_2,Q_2,U_2)$
is still $\Hom_S(M_1,M_2)$. (This point will be discussed further in
Section~\ref{morphisms} below.)
Then $\MFD_{\!\Pa}^{\,\epsilon}(W)$ is a differential graded category.
One can show that the linear map $P:\Hom_S(M_1,M_2)\to\Hom_S(M_2,M_1)$
in (\ref{Pdef})
commutes with the supercharge, since we have
$$
\forall\psi\in\Hom_S(M_2^*,M_1^*),\qquad
\dd (U_1\psi U_2^{-1})=U_1(\dd\psi)U_2^{-1}
$$ 
provided the condition (\ref{condi}) holds for both 
$(M_1,\sigma_1,Q_1,U_1)$ to $(M_2,\sigma_2,Q_2,U_2)$.
This means that we also have the homotopy category
$\MF_{\Pa}^{\,\epsilon}(W)$ consisting of invariant branes
with the constant $\epsilon$.

Similarly, we can construct the new categories based on the parity 
functor ${\mathcal P}=\aPa$.

\subsection{LG orbifolds}
\label{oo}

In orientifolding a LG orbifold, we consider the parity functor's
${\mathcal P}=\Pa_{\chi}(\tau)$'s introduced in (\ref{Pachi}).
An invariant brane is a quintuple $(M,\sigma,Q,\rho,U)$
where $(M,\sigma,Q,\rho)$ is an object of
$\MFD_{\Gamma}(W)$ and $U$ assigns to each odd element
$\tau\in \oGamma$ an isomorphism
\begin{equation}
U(\tau):M^*\to M
\end{equation}
such that
\begin{equation}
\begin{split}
&U(\tau)\sigma^T U(\tau)^{-1}
=\sigma,\\[0.1cm]
&U(\tau)(-\tau^*Q^T)U(\tau)^{-1}=Q,\\[0.1cm]
&U(\tau)(\chi\rho^{-T})U(\tau)^{-1}
=\rho.
\end{split}
\label{ccond}
\end{equation}
Without loss of generality, one can assume
\begin{equation}
U(g\tau)\rho(g)^T=\chi(g) U(\tau),\qquad
\forall g\in \Gamma,\quad
\forall \tau\in\oGamma\setminus\Gamma.
\eqlabel{see2}
\end{equation}
In fact, suppose we have $U(\tau)$'s obeying only (\ref{ccond})
but not necessarily (\ref{see2}).
Then, pick and fix any $\tau_0\in \oGamma\setminus \Gamma$ and modify
$U(\tau)$ by $\tilde{U}(\tau)=\rho(\tau\tau_0^{-1})U(\tau_0)$.
Then (\ref{see2}) as well as all the conditions in (\ref{ccond})
are satisfied for $\tilde{U}(\tau)$.
In what follows, we always assume the relation (\ref{see2}).

For two such invariant branes
$(M_i,\sigma_i,Q_i,\rho_i,U_i)$, $i=1,2$,
the parity transformation $\Hom_S(M_1,M_2)^{\Gamma}
\to \Hom_S(M_2,M_1)^{\Gamma}$ is defined as
\begin{equation}
P_{\chi}(\phi)=U_1(\tau)\tau^*\phi^TU_2(\tau)^{-1}.
\label{defPct}
\end{equation}
Because of the relation (\ref{see2})
this is independent of the choice of $\tau$.
At this stage, we require that $P_{\chi}$ be involutive on the equivariant
category. As before, the content of the requirement is found by computing
$P_{\chi}^2$, and yields the condition that
\begin{equation}
U(\tau)U(\tau)^{-T}\iota\rho(\tau^2)^{-1}=c(\tau)\cdot {\rm id}_M
\eqlabel{see}
\end{equation}
hold for any brane $(M,\sigma,Q,\rho,U)$, where $c(\tau)$
is a phase that is independent of the brane in a mutually compatible class.
Using $U(\tau)=c(\tau)\rho(\tau^2)\iota^{-1}U(\tau)^T$
twice, and with the help of the last condition in (\ref{ccond}), we find
the relation
\begin{equation}
c(\tau)^2\chi(\tau^2)=1.
\label{cchi}
\end{equation}
Combining (\ref{see}) and (\ref{see2}) and again with the help of
the last condition in (\ref{ccond}) we find
\begin{equation}
c(g\tau)=\chi(g)^{-1}c(\tau).
\label{cchic}
\end{equation}
This together with $c(\tau)^2\chi(\tau^2)=1$ means that the function 
$c(\tau)$ is, up to a sign, again uniquely determined by the choice
of parity.

Summarizing, we have defined the D-brane category
$\MFD_{\!\Pa_{\chi}}^{\pm c}(W)$ which is differential graded
and its homotopy category $\MF_{\!\Pa_{\chi}}^{\pm c}(W)$.

\subsection{Invariant branes from irreducible ones}

We have seen that in both LG model and LG orbifold,
the possible D-brane categories for a given parity are classified by
a sign: $\epsilon=\pm 1$ in (\ref{defep}) for the model without orbifold,
and for orbifold $c(\tau)$ in (\ref{see}) admits only two possibilities:
say, $c(\tau)=c_0(\tau)$ or $c(\tau)=-c_0(\tau)$.
These signs are actually the standard sign ambiguity for the
orientifold projection in the open string sector.

To see this, we study irreducible branes in the orbifold theory
and stacks of them
that can or cannot be members of one of these categories.
Let $B=(M,\sigma,Q,\rho)$ be an irreducible matrix factorization.
By definition, ``irreducible'' means that there is no subspace of $M$
that is invariant under both $\rho$ and $Q$.
Suppose it is also invariant under the parities
${\mathcal P}_{\chi}(\tau)$ for $\tau\in \oGamma$. Namely,
there is an $U(\tau):M^*\to M$ that obey (\ref{ccond}).
We deduce 
\begin{equation}
\begin{split}
Q(x) &= U(\tau) U(\tau)^{-T} \iota Q(\tau^2 x) \iota^{-1} U(\tau)^T 
 U(\tau)^{-1} \\
& = U(\tau) U(\tau)^{-T} \iota \rho(\tau^2)^{-1} Q(x) 
\rho({\tau^2}) \iota^{-1} U(\tau)^T U(\tau)^{-1} 
\end{split}
\end{equation}
{} From irreducibility, it follows
\begin{equation}
U(\tau) U(\tau)^{-T}\iota = \alpha(\tau) \rho(\tau^2)
\end{equation} 
where $\alpha(\tau)$ is a phase. Thence
$$
\alpha(\tau)^2\chi(\tau)^2=1.
$$
We also have $\alpha(g\tau)=\chi(g)^{-1}\alpha(\tau)$.
Therefore $\alpha(\tau)=\pm c_0(\tau)=:\epsilon^{\rm int}c_0(\tau)$, and
the ``internal'' sign $\epsilon^{\rm int}=\pm1$ is 
uniquely associated with $B$. We then choose an ``external'' Chan-Paton
space $V\cong \complex^N$ together with a map $\gamma: V^*\to V$ and define
\begin{equation}
\begin{split}
\hat M &= V \otimes  M  \\
\hat Q &= 1\otimes Q \\
\hat{U}(\tau) & = \gamma \otimes U(\tau) \qquad 
\tau\in\oGamma\setminus\Gamma \\
\hat \rho(g) & = 1\otimes \rho(g) \qquad g\in \Gamma 
\end{split}
\end{equation}
This definition satisfies the condition (\ref{ccond}), and we find
\begin{equation}
\hat{U}(\tau) \hat{U}(\tau)^{-T}\iota = \gamma\gamma^{-T} \otimes 
U(\tau) U(\tau)^{-T}\iota
 = \gamma\gamma^{-T} \otimes \alpha(\tau) \rho(\tau^2)
\end{equation}
Therefore, the brane
$(\hat{M},\hat{\sigma},\hat{Q},\hat{\rho},\hat{U})$
is a member of the category $\MFD_{\!\Pa_{\chi}}^{c}(W)$ with $c=\epsilon c_0$
if
$$\gamma\gamma^{-T}=
\epsilon^{\rm ext}=\epsilon^{\rm int}\epsilon.
$$
It is this sign, $\epsilon^{\rm ext}$, which decides 
whether the gauge group on $\hat Q$ is of symplectic or orthogonal 
type. This is as usual: Given an overall choice of orientifold sign
$\epsilon$, and the internal sign $\epsilon^{\rm int}$ which one 
{\it discovers} for any given $Q$ by an explicit computation, one 
can decorate this brane only with a CP space with appropriate 
$\epsilon^{\rm ext}$.

For completeness, we also consider the case where the irreducible brane
$B=(M,\sigma,Q,\rho)$ is not invariant, 
i.e. there is no $U(\tau):M^*\to M$ that obey (\ref{ccond})
and (\ref{see2}).
Then we can still form an invariant brane by combining
$B$ with its parity image and tensoring with an external Chan-Paton space
$V=\C^N$. Namely we define
\begin{equation}
\begin{split}
&\hat{M}=(V\otimes M)\,\oplus\,(V^*\otimes M^*),\\
&\hat{\sigma}=\left(\begin{array}{cc}
\sigma&0\\
0&\sigma^T
\end{array}\right),\\
&\hat{Q}=\left(\begin{array}{cc}
Q&0\\
0&-\tau_0^*Q^T
\end{array}\right)\\
&\hat{\rho}=\left(\begin{array}{cc}
\rho&0\\
0&\chi\rho^{-T}
\end{array}\right),
\end{split}
\label{invinv}
\end{equation}
where $\tau_0$ is an arbitrarily chosen odd element of the group $\oGamma$.
Then, this is invariant using $\hat{U}(\tau):\hat{M}^*\to\hat{M}$
given by
\begin{equation}
\hat{U}(\tau)=
\chi(\tau\tau_0^{-1})\left(\begin{array}{cc}
0&\alpha(\tau)\rho(\tau\tau_0)\iota^{-1}\\
\rho(\tau\tau_0^{-1})&0
\end{array}\right),
\label{noninvariant}
\end{equation}
where $\alpha$ is some phase.
If we choose $\alpha(\tau)=c(\tau)$,  this satisfies
\begin{eqnarray}
&&
\hat{U}(\tau)\hat{U}(\tau)^{-T}\hat{\iota}\hat{\rho}(\tau^2)^{-1}
=\left(\begin{array}{cc}
c(\tau)&0\\
0&c(\tau)
\end{array}\right),\nonumber\\
&&
\hat{U}(g\tau)\hat{\rho}(g)^T=\chi(g)\hat{U}(\tau).
\nonumber
\end{eqnarray}
Thus
$\hat{B}=(
\hat{M},\hat{\sigma},\hat{Q},\hat{\rho},\hat{U})$ is a member
of the category $\MFD_{\!\Pa_{\chi}}^{\,c}(W)$.
The gauge group is $U(N)$ in this case.

\subsection{Morphisms and gauge algebra}
\label{morphisms}

We have seen above that parity acts on the morphism spaces of both
$\MFD(W)$ and $\MF(W)$, and that requiring this action to be involutive 
leaves the freedom of an overall choice of sign, $\epsilon$, in the 
action of parity on individual objects. One is then tempted to try 
to impose parity invariance also on the morphism spaces. But some
care is required.

First note that even for invariant objects $B=(M,\sigma,Q,U)$ and 
$B'=(M',\sigma',Q',U')$, parity does 
not send the morphism space $\Hom_S(M,M')$ to itself, but to 
$\Hom_S(M',M)$ (where we are using the isomorphisms $U:M^*\to M$, 
$U':M'^*\to M'$). The minimal morphism spaces on which parity acts 
are, for $B\neq B'$, 
\begin{equation}
\oHom(M,M') := \Hom_S(M,M')\oplus \Hom_S(M',M)\,,
\eqlabel{minimal} 
\end{equation}
which can then indeed be decomposed into even and odd components under 
parity. For the {\it endomorphisms} of an invariant object $B=B'$,
we can be slightly more economical, and consider the action of parity
on
\begin{equation}
\oEnd(M) := \Hom_S(M,M)
\end{equation}
Note that $\oEnd(M)\neq \oHom(M,M)$ according to these definitions.

\begin{figure}[t]
\begin{center}
\psfrag{phi1}{$\phi_1$}
\psfrag{phi2}{$\phi_2$}
\psfrag{phi3}{$\phi_3$}
\psfrag{Pphi1}{$P(\phi_1)$}
\psfrag{Pphi2}{$P(\phi_2)$}
\psfrag{Pphi3}{$P(\phi_3)$}
\psfrag{P}{$P$}
\epsfig{height=5cm,file=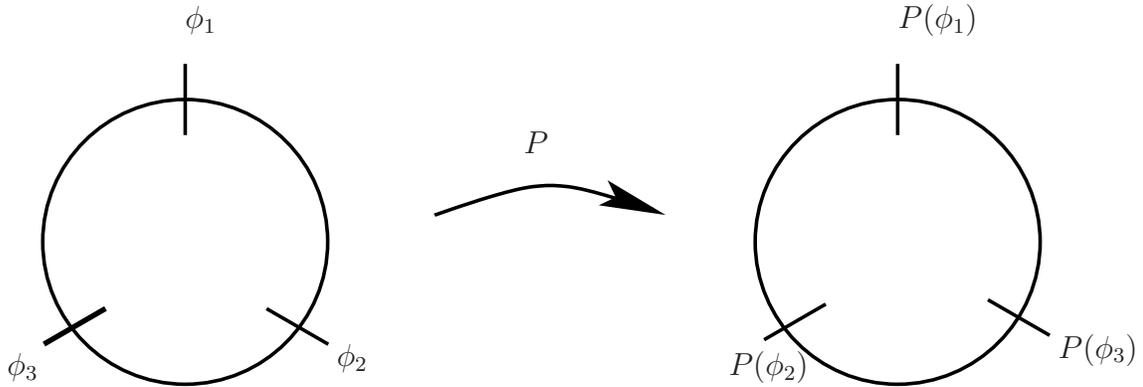}
\caption{A disk correlator and its parity image. The existence of this
symmetry {\it does not} (unfortunately) impose any restrictions on
the parities of $\phi_1$, $\phi_2$, $\phi_3$. For example, the correlator 
could be non-zero for $P_1=-,P_2=+,P_3=+$.}
\label{pardisk}
\end{center}
\end{figure}

The peculiarity of parity (as compared with other discrete symmetries)
is that it does not define an automorphism of the operator algebra
of open strings, but rather an {\it anti}-automorphism. Namely,
parity reverses the cyclic ordering of operators inserted on the
boundary of the worldsheet. See Fig.\ \ref{pardisk}. As a consequence,
parity {\it does not} impose any selection rule on worldsheet 
correlators with more than two boundary insertions. (For just
two insertions, which defines the topological metric, parity does 
yield a selection rule.) If these statements come as a surprise, we 
hasten to emphasize that of course in full string theory, orientifolding 
{\it does define} a projection that is consistent with string 
interactions. This is a consequence of integrating over the moduli 
space of Riemann surfaces, and in particular, summing over the ordering 
of operators on the boundary.

Nevertheless, even if the topological field theory does not admit
the ``orientifold projection'', we can still learn about the effect
of this projection in the string background built on the parity of
our Landau-Ginzburg model. Consider strings from one brane to itself. 
In both the ordinary and the orbifolded case, we have shown that
$P^2(\phi)=\phi$ for $\phi\in\Hom(M,M)$. We can then decompose
\begin{equation}
\oEnd(M) = \Hom(M,M) = \oEnd^+(M) \oplus \oEnd^-(M)
\end{equation}
where
\begin{equation}
P(A) = \pm A \qquad \text{for $A\in\oEnd^\pm(M)$, respectively.}
\end{equation}
As we have noted, we cannot consistently compose morphisms in a
way that is compatible with this decomposition. This is because
a diagram as Fig.\ \ref{pardisk} for $P(A)= P_A A$, $P(B)=P_B B$ 
only implies $P(AB)= (-1)^{|A||B|}P_A P_B BA$, which is not 
necessarily related to $AB$. However, the Lie algebra structure 
inherited from $\Hom(M,M)$,
\begin{equation}
\{A,B\} = AB -(-1)^{AB} BA
\end{equation}
is preserved up to a sign
\begin{equation}
P(\{A,B\}) = (-1)^{AB} \{P(B),P(A)\}
= - \{P(A),P(B)\}
\eqlabel{additional}
\end{equation}
where we are using that $P$ preserves the $\zet_2$ grading by
fermion number. Thus we see that, as a Lie algebra, we can indeed
decompose endomorphisms of an invariant object. In particular, the 
(degree zero component of the) {\it parity odd} part $\oEnd^-(M)$ will 
determine the gauge algebra on the brane worldvolume. It is easy to 
check that for invariant objects constructed from irreducible ones as 
in the previous subsection, the gauge algebra is indeed $su(N)$, 
$so(N)$, or $sp(N)$, depending on whether the object is invariant
or not, the internal sign $\epsilon^{\rm int}$, and the overall choice 
of $\epsilon$.

Turning now to the ``morphisms'' from $M$ to $M'$, we can decompose 
the spaces $\oHom(M,M')$ defined by \eqref{minimal} into even an odd
combinations under parity,
\begin{equation}
\begin{split}
\oHom(M,M')&= \oHom^+(M,M')  \oplus \oHom^-(M,M') \\
\oHom^\pm(M,M') &:= \bigl[ \Hom(M,M') \oplus \Hom(M',M)\bigr]^{P=\pm 1}
\end{split}
\eqlabel{decomp}
\end{equation}
Note that any $\hat\Phi\in \oHom(M,M')$ can be written as
\begin{equation}
\hat\Phi= (\Phi, \Phi^P) \qquad \text{with $\Phi^P= p_\Phi P(\Phi)$
for $\hat\Phi\in\oHom^{p_\Phi}(M,M')$, respectively.}
\end{equation}
This allows us to identify $\oHom^\pm(M,M')$ with $\Hom_S(M,M')$.
As noted above, although we can compose morphisms $\Phi\in\oHom(M,M')$ 
with morphisms $\Psi\in\oHom(M',M'')$, this
composition is not compatible with the decomposition into even/odd
under parity. However, it is easy to see that the parity is compatible
with the structure of $\oHom(M,M')$ as a super-bimodule over $\oEnd(M')
\times\oEnd(M)$ (as a Lie algebra).

Finally, let us observe that the decomposition \eqref{decomp} will 
become important to understand the triangulated structure of our
orientifold category: Clearly, only cones over invariant maps 
in $\oHom(M,M')$ will lead to invariant objects.

\subsection{(Extended) Kn\"orrer periodicity}

\def\Kn{{\Scr K}}
Kn\"orrer periodicity \cite{knoerrer} is the statement that the 
category, $\MF(W)$, of matrix factorizations for $W$ is equivalent 
to the category, $\MF(\tilde W)$ for $\tilde W = W+ xy$ with two 
additional variables. Kn\"orrer's equivalence $\Kn:\MF(W)\to
\MF(\tilde W)$
is given by
\begin{equation}
Q=\begin{pmatrix} 0 & f \\ g & 0 \end{pmatrix}
\quad\mapsto\quad
\Kn(Q) = \begin{pmatrix} 
0 & 0 & f & x \\ 
0 & 0 & -y & g \\
g & -x & 0 & 0 \\
y & f & 0 & 0
\end{pmatrix}
\eqlabel{horst}
\end{equation}
We want to understand how parity behaves under this equivalence.
What we mean by this is the relation between $\Pa$ and $\tilde \Pa$ 
in the diagram
\begin{equation}
\begin{matrix}
\MF(W) & 
\smash{\mathop{\longrightarrow}\limits^{\Pa}} &
\MF(W) \cr
\Big\downarrow\rlap{$\vcenter{\hbox{$\scriptstyle \Kn$}}$} & &
\Big\downarrow\rlap{$\vcenter{\hbox{$\scriptstyle \Kn$}}$} \cr
\MF(\tilde W) & 
\smash{\mathop{\longrightarrow}\limits^{\tilde \Pa}} &
\MF(\tilde W)
\end{matrix}
\end{equation}
We will here restrict ourselves to ordinary Landau-Ginzburg models. 
Generalization to orbifold case is straightforward.

At first, it seems that we have to fix a choice in the action on
the new variables $(x,y)$. We can have $(x,y)\mapsto (-x,y)$ or 
$(x,y)\mapsto (x,-y)$. However, the two possibilities differ
merely by conjugation by the global symmetry exchanging
$x$ and $y$, so the two parities should be considered equivalent.
Using conventions from \eqref{matrix}, \eqref{grtrans}, we have
\begin{equation}
\tilde \Pa(\Kn Q) = 
\begin{pmatrix}
0 & 0 & g^t & y \\
0 & 0 & x & f^t \\
-f^t & y & 0 & 0 \\
x & -g^t & 0 & 0
\end{pmatrix}
\end{equation}
{} from which we easily see
\begin{equation}
\Kn^{-1} \tilde \Pa(\Kn Q) \cong 
\begin{pmatrix}
0 & -f^t \\ g^t & 0
\end{pmatrix}
\end{equation}
Comparing this with
\begin{equation}
\Pa(Q) = 
\begin{pmatrix}
0 & g^t \\ -f^t & 0
\end{pmatrix}
\end{equation}
shows that $\tilde \Pa$ differs from $\Pa$ by an additional
orientation reversal, in other words, $\tilde \Pa \cong [1]\circ 
\Pa=\aPa$. Since $\Pa$ and $\aPa$ are {\it not} isomorphic 
parities, it is already clear that {\it Kn\"orrer periodicity
will be extended in the orientifold context}. In particular,
an invariant object in $\MF_\Pa^\epsilon(W)$ is mapped 
under $\Kn$ to an invariant object in $\MF_{\aPa}^\epsilon(W)$,
and except under exceptional circumstances, we do not expect
the two orientifold categories to be equivalent.

To make this more concrete, let us consider an invariant object 
in $\MF_\Pa^\epsilon(W)$ associated with the matrix factorization
$f\cdot g= W$ satisfying
\begin{equation}
\begin{pmatrix} 0 & f \\ g & 0 \end{pmatrix} =
\begin{pmatrix} U & 0 \\ 0 & V \end{pmatrix}
\begin{pmatrix} 0 & g^t \\ -f^t & 0 \end{pmatrix}
\begin{pmatrix} U^{-1} & 0 \\ 0 & V^{-1} \end{pmatrix}
\end{equation}
or $-U g^t V^{-1} = f$, $V f^t U^{-1}=g$. Being in 
$\MF_\Pa^\epsilon(W)$ means that
\begin{equation}
\begin{pmatrix}
U & 0 \\ 0 & V
\end{pmatrix}
\begin{pmatrix}
U & 0 \\ 0 & V
\end{pmatrix}^{-t}
= \epsilon
\begin{pmatrix} 1 & 0 \\ 0 & -1 \end{pmatrix}
\end{equation}
Let us see what happens in $\MF(\tilde W)$. We write (see
\eqref{horst})
\begin{equation}
\begin{split}
\tilde f = \begin{pmatrix} f & x \\ -y & g \end{pmatrix}
&\qquad
\tilde g = \begin{pmatrix} g & -x \\ y & f \end{pmatrix}  \\
\tilde U = \begin{pmatrix} 0 & U \\ V & 0 \end{pmatrix}
&\qquad
\tilde V = \begin{pmatrix} 0 & -V \\ U & 0 \end{pmatrix}
\end{split}
\end{equation}
and obtain the equivalence
\begin{equation}
\begin{pmatrix} 0 & \tilde f \\ \tilde g & 0 \end{pmatrix} =
\begin{pmatrix} 0 & \tilde U \\ \tilde V & 0 \end{pmatrix}
\begin{pmatrix} 0 & \tilde g^t \\ -\tilde f^t & 0 \end{pmatrix}
\begin{pmatrix} 0 & \tilde V^{-1} \\ \tilde U^{-1} & 0 \end{pmatrix} 
\end{equation}
or $\tilde f= -\tilde U \tilde f^t\tilde V^{-1}$,
$\tilde g = \tilde V \tilde g^t\tilde U^{-1}$.
This equivalence being odd is in contradiction to our definition
\eqref{condi}, and hence the mapped object is not in $\MF_\Pa^{\pm 
\epsilon}(W)$.

\def\ttilde#1{\tilde{\tilde #1}}
This extension of Kn\"orrer periodicity in the orientifold context
is reminiscent of the extension of periodicity in going from K-theory 
to real K-theory.\footnote{The extension of Kn\"orrer periodicity
in orientifolds was first noted in \cite{BHHW}, where it was given 
the more familiar (to physicists) interpretation as the distinction 
between type I string theory with D9/D5 branes and D7/D3 branes. This
is similar to ordinary Kn\"orrer periodicity as the distinction between 
type 0A and type 0B \cite{kali2}.} It is then natural to wonder how 
long is real Kn\"orrer 
periodicity. It is clear than when we iterate $\Kn$, the resulting parity
functor will again be $\Pa$. What remains to be checked is whether 
$\Kn^2$ maps $\MF_\Pa^\epsilon(W)$ to $\MF_\Pa^\epsilon(W)$ or to 
$\MF_\Pa^{-\epsilon}(W)$.

So let us check what happens when we iterate $\Kn$. We add another 
pair of variables, with $\ttilde \Pa:(u,v)\mapsto (-u,v)$, and
\begin{equation}
\ttilde Q = \begin{pmatrix} 0 &\ttilde f \\ \ttilde g & 0 \end{pmatrix}
\end{equation}
The invariance condition is 
\begin{equation}
\begin{pmatrix} 0 & \ttilde f \\ \ttilde g & 0 \end{pmatrix} =
\begin{pmatrix} \ttilde U & 0 \\ 0 & \ttilde V \end{pmatrix}
\begin{pmatrix} 0 & \ttilde g^t \\ -\ttilde f^t & 0 \end{pmatrix}
\begin{pmatrix} \ttilde U^{-1} & 0 \\ 0 & \ttilde V^{-1}  \end{pmatrix} 
\end{equation}
or $\ttilde f= \ttilde U \ttilde g^t\ttilde V^{-1}$,
$\ttilde g = -\ttilde V \ttilde f^t\ttilde U^{-1}$.
\begin{equation}
\begin{split}
\ttilde f = \begin{pmatrix} \tilde f & u \\ -v & \tilde g \end{pmatrix}
&\qquad
\ttilde g = \begin{pmatrix} \tilde g & -u \\ v & \tilde f \end{pmatrix}  \\
\ttilde U = \begin{pmatrix} 0 & \tilde U \\ \tilde V & 0 \end{pmatrix}
&\qquad
\ttilde V = \begin{pmatrix} 0 & -\tilde V \\ \tilde U & 0 \end{pmatrix}
\end{split}
\end{equation}
We now find
\begin{equation}
\ttilde\epsilon=\ttilde U\ttilde U^{-t}=\epsilon
\qquad
\ttilde\epsilon=\ttilde V \ttilde V^{-t} = -\epsilon
\end{equation}
So, $\ttilde Q$ has the same $\epsilon$ as $Q$. Kn\"orrer periodicity is 
just doubled.

\subsection{R-charge grading}
\label{rational}

We will now briefly discuss the compatibility of parity symmetries
with the additional $\rationals$ or $\zet$ gradings of the category
of matrix factorizations introduced in \cite{johannes}. We consider a 
homogeneous superpotential $W$. Namely, we assume that there exist
rational numbers $q_i$ such that with respect to the Euler vector
field $E=\sum q_i x_i\del/\del x_i$, we have
\begin{equation}
E W = 2 W
\end{equation}
$E$ provides the so-called R-charge grading of the Landau-Ginzburg
model. Let us consider a parity which leaves $E$ invariant when acting
on the $x_i$'s.

The R-charge grading of branes is discussed in detail in \cite{johannes}.
It is provided by an even matrix $R$ satisfying
\begin{equation}
EQ+[R,Q] = Q
\end{equation}
(with $EQ-Q$ measuring the obstruction to the existence of a grading). 
Clearly, under parity, $R$ transforms to its negative,
\begin{equation}
E Q^T + [P(R),Q^T] = Q^T
\end{equation}
with
\begin{equation}
P(R) = - R^T
\eqlabel{trivial}
\end{equation}
As also shown in \cite{johannes}, the RR charges of D-branes in 
the simplest class of Landau-Ginzburg orbifolds are essentially
determined by the R-charge grading of the brane. In particular,
by normalizing $R$ to $\tr R=0$, the so-called central charge of a brane
associated with $Q$ is given by the formula
\begin{equation}
Z(Q) = \str_M\, \sigma \ee^{\pi\ii (R-\varphi)} 
\end{equation}
where $\varphi$ is a phase such that $\bigl(\sigma\ee^{\pi\ii (R-\varphi)}
\bigr)^H=\id$. In this context, \eqref{trivial} shows that our LG parity
acts as a conjugation on the BPS charge lattice. This is in accord with 
general principles \cite{BH2,BHHW}.

It is also easy to see that parity always commutes with R-charge grading 
on open strings. Namely, for $\phi\in\Hom_S(M,M')$,
\begin{equation}
q_\phi \phi = E \phi + R'\phi - \phi R
\end{equation}
implies
\begin{equation}
q_{\phi} \phi^T = E\phi^T + P(R) \phi^T - \phi^T P(R')
\end{equation}
so $q_{P(\phi)}=q_\phi$.

\section{The Topological Crosscap State in Landau-Ginzburg Models}
\label{state}

In this section, we compute topological crosscap correlators in 
unoriented Landau-Ginzburg models. Our formula extends the
results of Vafa \cite{toplg} on closed string correlators and the
results of Kapustin-Li \cite{kali2}, see also \cite{hl}, for correlators 
on oriented surfaces with boundaries. We will also argue for the
extension of our result to orientifolds of Landau-Ginzburg orbifolds,
along the lines of \cite{johannes}. We will follow conventions of
those papers, and also refer to \cite{horietal} for general background
on the formulation of $\caln=2$ Landau-Ginzburg models and their
topological twist.

\subsection{Simple orientifold}
\label{simple}

\begin{figure}[ht]
\begin{center}
\psfrag{eq}{$=$}
\psfrag{B}{$B$}
\psfrag{C}{$C$}
\epsfig{width=\textwidth-2cm,file=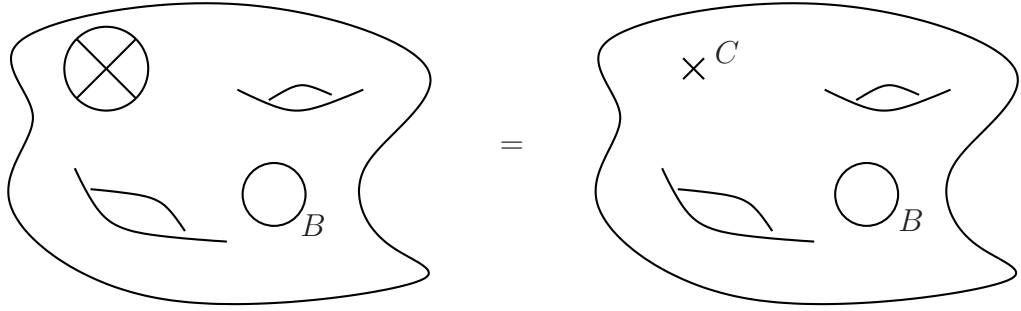}
\caption{A topological field theory correlator on a general unoriented
Riemann surface with boundary can be computed by replacing crosscaps 
(represented by a crossed circle) by the crosscap operator $C$. The 
open circle represents a boundary, with boundary condition labeled 
by $B$.}
\label{crossstate}
\end{center}
\end{figure}

Consider an involutive parity $P$ of a Landau-Ginzburg model defined as 
in \eqref{P} by a linear involution $\tau$,
\begin{equation}
x \mapsto \tau(x), \qquad \tau(x)^i = \tau^i_j x^j
\end{equation}
satisfying
$W(\tau x) = \tau(W(x)) = -W(x)$ and $\tau^2=1$. We wish to compute the 
topological crosscap correlator
\begin{equation}
\langle \phi \rangle_{C}
\eqlabel{corr}
\end{equation}
where $\phi$ is an arbitrary bulk insertion, namely, an element of
the Jacobi ring
\begin{equation}
\phi\in\calj_W = \complex[x_1,\ldots,x_r]/\del W
\end{equation}
Knowledge of these correlators, together with non-degeneracy of the closed 
string topological metric, will allow us to write down the ``crosscap operator'', 
$C$, and therefore the correlators of topological field theory on a general 
unoriented Riemann surface with arbitrary numbers of boundaries and handles,
as illustrated in Fig.\ \ref{crossstate}.

By definition, the correlator \eqref{corr} is given by the path-integral
on the $\reals\projective^2$ worldsheet with Lagrangian
\begin{equation}
|\del x^i|^2 + |\del_i W|^2 +
\psi^{\bar i} \delbar \rho_z^i + \bar\psi^{\bar i} \del \rho_{\bar z}^i +
\frac 12 \del_i\del_j W \rho_{\bar z}^i \rho_z^j +
\frac 12 \del_{\bar i}\del_{\bar j} \bar W \psi^{\bar i} \bar \psi^{\bar j}
\end{equation}
Here, as usual (see, \eg, \cite{horietal}), $\rho_z^i$/$\psi^{\bar i}$ are 
complex one-forms/scalars valued in the holomorphic/anti-holomorphic tangent 
bundle, respectively.

We can view the path-integral on $\reals\projective^2$ as the path-integral
over the subset of fields on the sphere worldsheet $\complex\projective^1$, 
parametrized by $z$, and subject to the crosscap boundary conditions
\begin{equation}
\begin{split}
x^i(-1/\bar z) &= \tau^i_j x^j(z) \\
\psi^{\bar i}(-1/\bar z) &= \tau^{\bar i}_{\bar j} \bar \psi^{\bar j}(z) \qquad
\bar \psi^{\bar i}(-1/\bar z) = \tau^{\bar i}_{\bar j} \psi^{\bar j}(z) \\
\rho_z^i(-1/\bar z) &= \tau^i_j \rho_{\bar z}^j (z) \qquad
\rho_{\bar z}^i (-1/\bar z) = \tau^i_j \rho_z^j (z) 
\end{split}
\end{equation}
where $\tau^{\bar i}_{\bar j} = \bar \tau^i_j$ (but it's no loss of generality
to assume that $\tau$ is real).

As usual, the path-integral can be localized to the zero modes,
which means constant $x=\tau(x)$ and constant $\psi = \tau(\bar\psi)$, 
$\rho=\tau(\bar\rho)$. In other words, we are restricted to the computation 
of the finite-dimensional integral
\begin{equation}
\int dx\; d\bar x\; d\psi\; d\rho\; 
\exp\Bigl(-|\del W|^2 + \frac 12 \del_i\del_j W \rho_{\bar z}^i \rho_z^j
+ \frac 12\del_{\bar i}\del_{\bar j} \bar W \psi^{\bar i} \bar\psi^{\bar j}
\Bigr)
\eqlabel{finite}
\end{equation}
over the invariant part of the target space,
\begin{equation}
 x= \tau x\qquad \psi^{\bar i} = \tau^{\bar i}_{\bar j} \psi^{\bar j}
\qquad \rho^i_z = \tau^i_j \rho^j_{\bar z}
\end{equation}
We expect that when the critical points of $W$ are isolated, the 
integral \eqref{finite} localizes further. Namely, we expect a sum
over critical points of $W$, the contribution of each of which
is obtained in the linear approximation. This is how the computation
was done in \cite{toplg} in the oriented case, and in \cite{kali2,hl} 
in the presence of boundaries. Namely, first the superpotential is
resolved by addition of relevant deformations, and then the result of
that computation is continued to the degenerate case. To proceed 
along these lines in the present situation, we should deform the 
superpotential in such a way as to preserve the condition
$\tau(W)=-W$. It is reasonable to assume that there always exists such
a deformation which completely resolves the superpotential.

Let us then assume that the critical points of $W$ are isolated. Any given
critical point can be invariant under $\tau$ or it can be mapped to another
critical point. Only the invariant critical points contribute to \eqref{finite}.
Let us consider the contribution from one of them. We note that 
from $\del_i W (x) = - \del_i \bigl(W(\tau x)\bigr) = -(\del_j W)(\tau x) 
\tau^j_i$, it follows that at an invariant critical point, 
$|\del W|^2$= $|\del^\perp W|^2$ where $\del^\perp$ is the derivative 
in the direction perpendicular to the fixed locus of $\tau$. Actually, let
us introduce coordinates $x_\perp^i$ and $x_\parallel^i$ which are
invariant and anti-invariant under $\tau$, respectively, 
$\tau^i_j x_\parallel^j = x_\parallel^i$, $\tau^i_j x_\perp^j= -x_\perp^i$.
Since $\tau(W)=-W$, we have
\begin{equation}
W(x) = x_\perp^i R_i(x_\perp,x_\parallel)
\eqlabel{factorize}
\end{equation}
for some choice of polynomials $R_i$ which are invariant under $\tau$.
Note that the $R_i$ are not uniquely determined by this condition. Eq.\ 
\eqref{factorize} can be used to define a matrix factorization of $W$.
The upshot of our computation will be that the crosscap state can be
identified with the boundary state \cite{kali2} corresponding to this
particular factorization of $W$.

Let us first study the contribution from the integration over
fermionic zero modes. Because of topological twisting, there is only 
one $\psi$ zero mode, and no $\rho$ zero mode. Since $\bar \psi^{\bar j} 
= \tau^{\bar j}_{\bar k} \psi^{\bar k}$, the integral is
\begin{equation}
\int d\psi \exp 
\Bigl(\frac 12 \del_{\bar i} \del_{\bar j} \bar W \psi^{\bar i}
\tau^{\bar j}_{\bar k} \bar\psi^{\bar k} \Bigr)
= \Pf \bigl(\del_{\bar i}\del_{\bar k} \bar W \tau^{\bar k}_{\bar j}\bigr)
\end{equation}
The appearance of the Pfaffian is of course not unexpected. It
makes sense since $\del_i\del_k W (x) \tau^k_j = - \del_i \bigl((\del_j 
W)(\tau x)\bigr) = -(\del_k \del_j W)(\tau x) \tau^k_i$, and therefore, at an 
invariant critical point, $\del_i\del_k W \tau^k_j$ is antisymmetric in 
$i$ and $j$. The sign of the Pfaffian is related to the overall choice 
of the crosscap.

The integral over bosonic zero modes gives in the linear approximation
\begin{equation}
\int dx_{\parallel} d\bar x_\parallel \exp\bigl(-|\del^\perp W|^2\bigr)
= \frac{1}{\det \del^\parallel_i\del^\perp_k W \del^\parallel_{\bar j} 
\del_\perp^{\bar k} \bar W}
\end{equation}

Our result for the topological crosscap correlator is therefore first
given by a sum over invariant critical points of $W$
\begin{equation}
\langle \phi \rangle_C = \sum_{\topa{x=\tau x}{\del W=0}} \phi\;
\frac{\Pf \bigl(\del_{\bar i}\del_{\bar k} \bar W \tau^{\bar k}_{\bar j}\bigr)}
{\det \bigl(\del^\parallel_i\del^\perp_k W \del^\parallel_{\bar j} 
\del_\perp^{\bar k} \bar W\bigr)}
\eqlabel{preliminary}
\end{equation}
This formula becomes more transparent if we use that with respect to the 
$(x_\parallel^i,x_\perp^i)$ coordinates introduced above, the matrix of second
derivatives of $W$ is block off-diagonal (at an invariant critical point),
\begin{equation}
\HH = \bigl(\del_i\del_j W\bigr)_{i,j} =
\begin{pmatrix} 0 & B \\ B^T & 0 \end{pmatrix}
\end{equation}
where $B$ is the matrix $B_{ij} = \del^\parallel_i\del^\perp_j W$.
In these coordinates, it is easy to see that $\HH$ is non-degenerate
only if the number of $x_\perp^i$ is equal to the number of $x_\parallel^i$, 
\ie, $B$ is a square matrix. In particular, the number of variables, $r$,
must be even. Then, $\Pf\bigl(\del_{\bar i}\del_{\bar k} \bar W \tau^{\bar 
k}_{\bar j}\bigr)=(-1)^{r/2} \det \bar B$, and 
$\det \bigl(\del^\parallel_i\del^\perp_k W \del^\parallel_{\bar j} 
\del_\perp^{\bar k} \bar W\bigr) = |\det B |^2$. Moreover, the Hessian 
is $H=\det\HH= \det \del_i\del_j W = (\det B)^2$ and at an invariant
critical point,
\begin{equation}
\frac{(-1)^{r/2}\det \bar B}{|\det B|^2}
= \frac{1}{\Pf \HH \tau} = \frac{\Pf \HH \tau}{H}
\end{equation}

We are now in a position to derive the crosscap state corresponding to
the parity $\tau$. First of all, when the number of variables is odd, or
$\tr \tau\neq 0$, we will simply have $C=0$. Any complete resolution of 
the singularity will not have any invariant critical points. When $\tr 
\tau=0$, (the number of $x^i_\parallel$ is equal to the number of 
$x^i_\perp$), we claim that the crosscap state, $C$, is (up to a sign) 
nothing but the boundary state of \cite{kali2} corresponding to the 
factorization, \eqref{factorize}, namely
\begin{equation}
C \; dx_\parallel^1\wedge dx^2_\parallel\wedge\ldots\wedge dx_\perp^{r/2} = 
\frac{(-1)^{\frac r2\left(\frac r2+1\right)/2}}{r!} 
\str (\del Q)^{\wedge r} \,,
\eqlabel{crosscap}
\end{equation}
where $Q$ is the odd matrix 
\begin{equation}
Q = \sum_{i=1}^{r/2} \bigl(x^i_\perp \pi_i + R_i \bar\pi^i\bigr)
\eqlabel{Q}
\end{equation}
and $(\pi_i,\bar\pi^i)$ generate a Clifford algebra, $\{\pi_i,\bar\pi^j\}=
\delta_i^j$. More explicitly, the claim is
\begin{multline}
C\; dx^1_\parallel\wedge\ldots \wedge dx^{r/2}_\perp 
= \frac{(-1)^{\frac r2\left(\frac r2+1\right)/2}}{r!} 
\str \Bigl({\textstyle\sum_{i=1}^{r/2}}\; 
\bigl(dx^i_\perp \pi_i + d R_i\bar\pi^i
\bigr)\Bigr)^{\wedge r} \\
= (-1)^{r/2} 
\epsilon^{i_1i_2\ldots i_{r/2}} \bigl({\textstyle
\prod_{s=1}^{r/2}} \; \del_{i_s}^\parallel R_{s}\bigr)
dx^1_\parallel\wedge \ldots dx^{r/2}_\parallel \wedge dx^1_\perp
\wedge\ldots dx^{r/2}_\perp
\end{multline}
where we have used that $\str \,\pi_i\bar\pi^j=\delta_i^j$. In other words
\begin{equation}
C = (-1)^{r/2}
\epsilon^{i_1i_2\ldots i_{r/2}} \bigl({\textstyle
\prod_{s=1}^{r/2}} \; \del_{i_s}^\parallel R_{s}\bigr)
\eqlabel{explicit}
\end{equation}

\begin{figure}[t]
\begin{center}
\psfrag{eq}{$=$}
\psfrag{phi}{$\phi$}
\psfrag{C}{$C$}
\epsfig{width=\textwidth-4cm,file=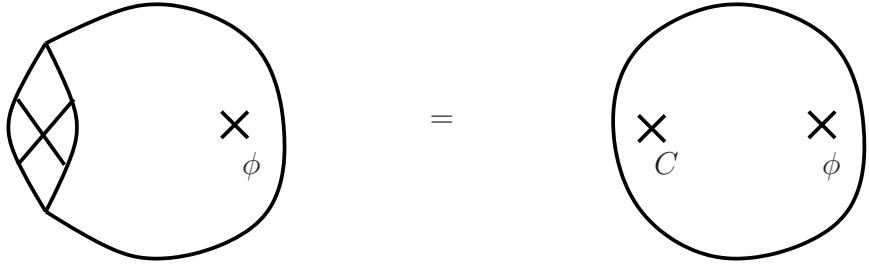}
\caption{The crosscap correlator we have computed is equivalent to
a sphere correlator with insertion of the crosscap operator $C$.}
\label{cst}
\end{center}
\end{figure}

To prove this claim, we have to show that for all $\phi\in\calj_W$,
\begin{equation}
\langle \phi\rangle_C = \langle C\phi\rangle_0
\end{equation}
where the right hand side is a sphere correlator, see Fig.\ \ref{cst}.
According to \cite{toplg}, the sphere correlator is given by
\begin{equation}
\langle C \phi\rangle_0 = \sum_{\del W=0} \frac{C\phi}{H}
\end{equation}
Here the sum is over all critical points of $W$ (as opposed to only 
the invariant ones), so the claim follows if it is true that
\eqref{explicit} coincides with $\Pf\HH \tau= (-1)^{r/2}\det B$ at an 
invariant critical point, and vanishes at a non-invariant one. The 
first assertion derives from \eqref{factorize}, since
\begin{equation}
B_{ij} = \del_i^\parallel\del_j^\perp W|_{x_\perp=0} = \del^\parallel_i R_j
\end{equation}
so $C=(-1)^{r/2} \det B$. On the other hand, half of the $r$ conditions for 
a non-invariant critical point, $0 = \del_j^\parallel W = x^i_\perp 
\del_j^\parallel R_i$ with $x^i_\perp$ not all zero, imply that 
$\del_j^\parallel R_i$ is a degenerate matrix, and therefore its 
determinant is zero. The claim follows.

As we have mentioned, if $W$ is degenerate, we should first deform 
it such that parity is respected and such that all the critical points 
of $W$ are isolated. By continuity, the formula \eqref{crosscap} will 
then hold also in the degenerate case. We now proceed to check this in 
examples.

\subsection{Examples}
\label{examples}

Consider a D-type minimal model
$$
W = x^{k/2+1} + x y^2
$$
with $k/2$ even. (We do not consider the models with $k/2$ odd because
they do not admit an involutive parity.) There are two choices of parity 
$P_{\pm}$, associated with the action $\tau_{\pm}: (x,y)\mapsto (-x,\pm y)$
on the chiral fields. Clearly, the crosscap state 
for $\tau_-$ is simply zero, $C_-=0$. The crosscap for $\tau_+$ is 
represented by the boundary state for
\begin{equation}
Q = \begin{pmatrix} 0 & x \\ x^{k/2} + y^2 \end{pmatrix}
\end{equation}
\ie,
\begin{equation}
C_+ dx\wedge dy = \frac 12 \bigl(\str\, dQ^{\wedge 2}\bigr) =  2y\,dx\wedge dy
\end{equation}
or $C_+=2y$. The simplest check that this is correct comes from
the Klein bottle amplitude, which computes $\tr (-1)^F\tau$
in the closed string sector. Representatives of the chiral 
ring are $(1,x,\ldots, x^{k/2-1},y, y^2 = - \bigl(\frac 
k2+1\bigr) x^{k/2})$. We see that
\begin{equation}
\tr(-1)^F = \frac k2+2 = 
\langle H \rangle_0 =  \left\langle k \bigl({\textstyle\frac k2}+1\bigr) x^{k/2}
- 4 y^2\right\rangle_0 = -(k+4)\langle y^2\rangle_0
\end{equation}
which shows that the correct normalization of the sphere correlator is
$\langle y^2\rangle_0=-1/2$. 
\begin{figure}[t]
\begin{center}
\psfrag{eq}{$=$}
\psfrag{C}{$C$}
\epsfig{width=\textwidth-4cm,file=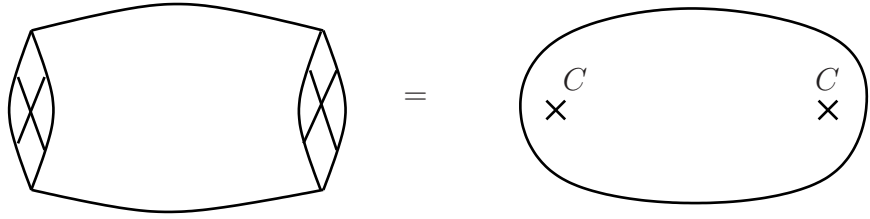}
\caption{The Klein bottle correlator can be expressed in terms of the square
of the crosscap operator on the sphere.}
\label{kleincorr}
\end{center}
\end{figure}
The Klein bottle being representable as two 
crosscaps connected by a cylinder (see Fig.\ \ref{kleincorr}), general 
principles of TFT demand that the Klein bottle amplitude for the parity 
$P_+$ be given by
\begin{equation}
K_+ = \tr(-1)^F P_+ = \langle C_+^2 \rangle_0 = \langle 4 y^2\rangle_0
= -2
\end{equation}
This is (up to a sign) indeed the correct result based on the above 
representation of the chiral ring, which has $k/4+2$ even and $k/4$ odd 
elements,
\begin{equation}
K_+ = \tr(-1)^F P_+ = - \tr_{\calj_W} \tau_+
\eqlabel{klein}
\end{equation}
Note the (model-independent) sign difference between $\tr(1-)^F P$ over 
the RR ground states and $\tr \tau$ over the chiral ring.
The analog of equation \eqref{klein} also holds for $P_-$, where $C_-=0$, 
and the chiral ring has $(k/2+1)/2$ even and $(k/2+1)/2$ odd generators.

As another example, we consider
\begin{equation}
W = x^5-y^5
\end{equation}
with parity $\tau:(x,y)\mapsto (y,x)$. Our crosscap is easily computed
to be
\begin{equation}
C = 5 (x^3+ x^2y+ xy^2+y^3) 
\end{equation}
Since the dimension of the chiral ring is $16=-(5\cdot 4)^2\langle x^3y^3 
\rangle_0$, we find $\tr (-1)^F \tau = \langle C^2\rangle_0 = -4$, which 
coincides with $-\tr \tau$ over the chiral ring, as it should be.

\begin{figure}[t]
\begin{center}
\psfrag{B}{$B$}
\psfrag{C}{$C$}
\psfrag{eq}{$=$}
\epsfig{width=\textwidth-4cm,file=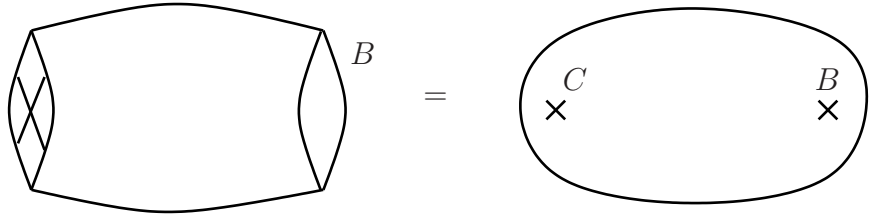}
\caption{The M\"obius correlator can also be computed using the
crosscap and boundary states.}
\label{mobiuscorr}
\end{center}
\end{figure}

We have also checked in some examples that the M\"obius correlator with 
boundary condition $B$ correctly gives the trace of $(-1)^F P$ acting in 
the Hilbert space of open strings $\calh_{B,\Pa(B)}$ from $B$ to its parity 
image $\Pa(B)$,
\begin{equation}
M_B = \tr_{\calh_{B,\Pa(B)}} (-1)^F P = \langle C B \rangle_0
\eqlabel{mcorr}
\end{equation}
where $B$ is the ``boundary state'' of \cite{kali2} associated with the boundary
condition of the same name.
See also Fig.\ \ref{mobiuscorr}.

\subsection{Charges and index theorem in general Landau-Ginzburg
orientifolds}

We now turn to the general case of Landau-Ginzburg orientifold,
in which the orientifold group is an extension of the form
\eqref{origroup}.

We wish to compute the parity twisted index
$\tr (-1)^F P$ in the open string sector between a matrix
factorization $(M,\sigma,Q,\rho)$ and its parity image.
We recall that a parity sends a brane based on a module $M$ 
to a brane based on the dual $M^*$.
Thus we are considering a parity operator that maps 
$\Hom_S(M,M^*)$ to $\Hom_S(M^{**},M^*)$.
But in order to take the trace, we need to have an operator that acts on
the same space. In other words, we need to make a choice of
relating $M^{**}$ to $M$.
The canonical one $\iota:M\to M^{**}$ is defined in (\ref{iota}),
but this is just one choice and one has to check that it is
consistent with everything.

This difficulty is cleanly solved if we use one of the
categories introduced in Section~\ref{category}.
Let us consider the category $\MFD_{\Pa_{\chi}}^{c}(W)$
where $\Pa_{\chi}$ is the parity functor defined in (\ref{Pachi})
and $c$ is the function $c(\tau)$ obeying
the equation
(\ref{cchi}) and (\ref{cchic}).
Then, we consider the invariant brane 
$(\hat{M},\hat{\sigma},\hat{Q},\hat{\rho},\hat{U})$ in this
category that includes
$M$ as a part, as $\hat{M}=M\oplus M^*$.
Thus, we take $V=\C$ ($N=1$) and we set $\alpha(\tau)=c(\tau)$ in
(\ref{invinv}) and
(\ref{noninvariant}).
Then the parity operator $P_{\chi}$ maps $\Hom_S(\hat{M},\hat{M})$ to 
itself and sends the subspace
$$
\Hom_S(M,M^*)\subset \Hom_S(\hat{M},\hat{M})
$$
to itself. Then there is no ambiguity in defining the trace.

Concretely, the requisite isomorphism is given by $\rho(\tau^2)c(\tau)
\iota^{-1}:M^{**}\to M$, where $\iota$ is the canonical isomorphism 
\eqref{iota}. With these definitions, the parity image (in $\Hom_S(M,M^*)$)
of a morphism $\Phi\in\Hom_S(M,M^*)$ is
\begin{equation}
P(\Phi)(x) =  \Phi^T(\tau x) \iota c(\tau)^{-1}
\rho(\tau^2)^{-1}
\end{equation}
To compute the index $\tr(-1)^F P$, we proceed as in 
\cite{johannes}. Namely, we choose a regularization
such that we can compute the index on the complex instead
of on the cohomology. A useful basis for the morphisms
in $\Hom_\calr (M,M^*)$ is
\begin{equation}
\Phi(x) = \sum_{ij,\alpha} \Phi^{ij}_\alpha e_{ij} x^\alpha
\eqlabel{basis}
\end{equation}
where $\alpha=(\alpha_1,\ldots,\alpha_r)$ is a multi-index. Let us
assume, as in section \ref{rational}, that $\tau$ is a phase rotation 
combined with at most an order two permutation of the variables,
namely, $\tau(x^\alpha)=\tau_\alpha x^{\tau(\alpha)}$, where
$\tau_\alpha$ is a phase. Parity then acts on basis elements as 
$P(e_{ij} x^\alpha)= e_{ji}\sigma^{i+j} \tau_\alpha x^{\tau(\alpha)}$.
(See section \ref{conventions} and in particular eqs.\ \eqref{matrix}
and \eqref{grtrans} for conventions.)

Taking the trace over morphism space requires summing over $i=j$ 
and $\alpha=\tau(\alpha)$. In this sector, 
$(-1)^F=1$ and hence we have
\begin{equation}
\tr (-1)^F P  = \sum_{j} \sigma_j c(\tau)^{-1}
\rho(\tau^2)^{-1}_{jj} \sum_{\alpha=\tau(\alpha)} \tau_\alpha
\end{equation}
where $\sigma_j$ originates from the fact that in the matrix 
representation, \eqref{dualdual}, $\iota$ can be identified
with $\sigma$.

Generally, we can separate the variables into those, $x^u_1,x^u_2,
\ldots ,x^u_{r_\tau}$ with $\tau_i^2=1$ (untwisted ones) and the
remaining ones with $\tau_i^2\neq 1$ (twisted ones). Let us first 
assume that there are no untwisted variables. Actually, let us 
assume that the action of $\tau$ has been diagonalized. The result 
is then simply
\begin{equation}
\tr (-1)^F P = \str \rho(\tau^2)^{-1} \frac{c(\tau)^{-1}}
{\prod_i (1-\tau_i)}
\end{equation}
When there are untwisted variables, $\tr(-1)^F P$ can be computed 
by combining the present method with the results from subsection 
\ref{simple}, see eq.\ \eqref{mcorr}. This is again similar to 
\cite{johannes}. Namely, we can contemplate 
an ``effective'' Landau-Ginzburg model with superpotential 
$W_\tau$ obtained from $W$ by setting all twisted variables to 
zero. If we also set the twisted variables to zero in $Q$,
we obtain an effective factorization $Q_\tau$ of $W_\tau$. The 
parity acting in this sector squares to $1$, so the results of 
subsection \ref{simple} are applicable. Defining $C_\tau$ for 
example by the formula \eqref{explicit} (or by zero when $\tr 
\tau\neq 0$ in this sector), the index formula takes the form
\begin{equation}
\begin{split}
\tr (-1)^F P &= 
\frac{c(\tau)^{-1}}{\prod_{i,\tau_i^2\neq 1}(1-\tau_i)}
\frac 1{r_\tau!} \res_{W_\tau} 
\bigl( \phi^\alpha_\tau 
\str\bigl[\rho(\tau^2)^{-1} (\del Q_\tau)^{\wedge r_\tau}\bigr]\bigr) \;
\eta_\tau^{\alpha\beta} \; \res_{W_\tau} \bigl(\phi^{\beta^*}_\tau \;C_\tau\bigr) \\
&= 
\frac{c(\tau)^{-1}}{\prod_{i,\tau_i^2\neq 1}(1-\tau_i)}
\frac{1}{r_\tau!} \res_{W_\tau} \bigl( 
\str\bigl[\rho(\tau^2)^{-1} (\del Q_\tau)^{\wedge r_\tau}\bigr]\;
C_\tau \bigr)
\end{split}
\eqlabel{simpleindex}
\end{equation}
where $r_\tau$ is the number of untwisted variables, $(\phi^\alpha_\tau)$
is a basis of the half-charged chiral ring $\calj_\tau=\complex[x_1^u,
\ldots,x_{r_\tau}^u]/\del W_\tau$ with R-charge $q(\Phi^\alpha_\tau)=\hat 
c_\tau/2$, and $\eta^{\alpha\beta}_\tau$ is the inverse of the closed
string topological metric in this sector. Finally, $\res_{W_\tau}$
is the residue that appears in closed string topological correlators
\cite{toplg}
\begin{equation}
\res_{W_\tau} f = \oint \frac{f}{\del_1 W_\tau\cdots\del_{r_\tau}W_\tau}
\end{equation}
As we have encountered it in section \ref{examples}, $\res_{W_\tau}$ 
is normalized in such a way that $\res_{W_\tau} H_\tau$
is equal to the dimension of the chiral ring of $W_\tau$. ($H_\tau$
is the Hessian of $W_\tau$.)

Using these formulas, it is a simple matter to implement the
orbifold projection on $\tr(-1)^F P$, by summing over $g\in\Gamma$
\begin{equation}
\frac{1}{\Gamma} \sum_{g\in\Gamma} \tr (-1)^F gP
\end{equation}

For example, when $\Gamma\cong\zet_H$ itself is cyclic and generated
by $g$, we have with obvious notation
\begin{equation}
\begin{split}
\tr (-1)^F P &= \frac 1H \sum_l 
\frac{c(g^{l} \tau)^{-1}}
{\prod_{i,\tau_i^{2l}\neq 1} (1-g_i^l\tau_i)}
\frac{1}{r_l!}\res_{W_{l}} \bigl(\str\bigl[\rho(g^{2l}\tau^2)^{-1}\;
(\del Q_l)^{\wedge r_l}\bigr] \; C_l \bigr)  \\
= \frac 1H \sum_l &
\frac{\chi(g^l) c(\tau)^{-1}}
{\prod_{i,\tau_i^{2l}\neq 1} (1-g_i^{2l}\tau_i^{l})}
\frac{1}{r_l!}\res_{W_{l}} \bigl(\str\bigl[\rho(g^{2l}\tau^2)^{-1}\;
(\del Q_l)^{\wedge r_l}\bigr] \; C_l \bigr)  
\prod_{i,\tau_i^{2l}\neq 1} (1+g_i^l\tau_i)
\end{split}
\eqlabel{indextheorem}
\end{equation}
Using these index formulae, we can derive the expression for
the RR charge of the crosscap. We recall \cite{toplg} that the 
relevant Ramond ground states are labeled by $|l;\alpha\rangle$,
where $l$ labels 
twisted sector, and $\alpha$ runs over a basis of the half-charged
chiral ring $\calj_l$. The first thing to notice is that
the Ramond charge is zero when there is no parity in $\oGamma$
that squares to $g^l$. When there is, we have a sum over parities 
whose square is $g^l$. Namely,
\begin{equation}
\langle l;\alpha \rangle_C = 
\sum_{\tau,\tau^2=g^l} \chi(g^l)c(\tau)^{-1} 
\;\bigl({\textstyle\prod_{i,\tau_i^{2l}\neq 1}} 
(1+\tau_i)\bigr)\;
\res_{W_l} \bigl( \phi^\alpha_l C_l\bigr)
\end{equation}
Indeed, recalling from \cite{johannes} the expression for the
D-brane charge
\begin{equation}
 \langle l;\alpha|Q\rangle_{\rm disk} =
\frac{1}{r_l!}\res_{W_l} \bigl( \phi_l^\alpha\; \str\bigl[\rho(g^l) 
(\del Q_l)^{\wedge r_l}\bigr] \bigr)
\end{equation}
we see that the index theorem \eqref{indextheorem} can be written as
\begin{equation}
\tr(-1)^F P 
= \frac1H\sum_{l=0}^{H-1} \sum_{\alpha,\beta}
\langle l;\alpha|Q \rangle_{\rm disk}
\frac{1}{\prod_{lq_i\notin 2\zet} (1-g_i^l)}
\eta_l^{\alpha\beta}
\langle l;\beta\rangle_C^*
\end{equation}

\section{Concluding Remarks}
\label{final}

In this paper, we have studied parity functors and constructed 
associated orientifold categories of D-branes in a specific class of 
backgrounds, Landau-Ginzburg models. However, the basic procedure found 
in this paper can be applied more generally. 
See the beginning of Section~\ref{category}.
We first choose a parity, an anti-involution of
the category of D-branes.
This allows us to consider invariant objects;
an invariant object is a brane whose parity image
is isomorphic to itself where the isomorphism is included as a
part of the data.
For each pair of invariant objects,
we can also consider a parity operator
whose square is an automorphism of the space of states.
We then classify invariant objects in such a way that
the parity operators are involutive for the members of one class.
Invariant objects of each class form an orientifold category.
We decide to keep all morphisms of the original category as
the morphism of the orientifold category.
What we found in Landau-Ginzburg models is that
there are exactly two classes, hence two categories, for each
parity functor.

This is reminiscent of {\it Hermitian K-theory}\footnote{E. Getzler pointed
out to us the relevance of Hermitian K-theory after the talk \cite{talk}.
KH thanks Max Karoubi for instruction. See the introduction of
\cite{Karoubi}, and also \cite{Mishchenko} for a survey.}
which can be defined for an algebra $A$ over some field $k$
with an anti-involution (an anti-involution
 is a $k$-linear automorphism of $A$, $a\mapsto \overline{a}$,
such that $\overline{ab}=\overline{b}\overline{a}$).
The construction goes as follows.
Let $M$ be a right $A$-module.
Its dual $M^t$ is defined as the set of all
$k$-linear maps $\sigma:M\to A$ such that $\sigma(ma)=\overline{a}\sigma(m)$,
which is again a right $A$-module.
The double dual is canonically isomorphic to the original,
$\iota:(M^t)^t\to M$, $\iota(m)(\sigma):=\overline{\sigma(m)}$.
Then, Hermitian K-theory ${\rm KQ}^{\varepsilon}(A)$ is the
Grothendieck group of the pair $(M,U)$ where
$U:M^t\to M$ is an isomorphism of $A$-modules such that
$$
U(U^t)^{-1}\iota=\varepsilon.
$$
Again, there is a sign ambiguity for $\varepsilon$.
If we take as $A$ the (commutative)
algebra over $\C$
of continuous complex-valued functions of a topological space $X$
with an involution $\tau$, with $\overline{f}(x):=f(\tau x)$,
then the Hermitian K-theories with $\varepsilon=\pm1 $ are nothing 
but Real K-theories ${\rm KR}(X),$ ${\rm KR}^{-4}(X)$
of Atiyah \cite{atiyah}. Since KR-theory classifies D-brane charges 
in geometric orientifolds \cite{realK}, it is clear that the structure
contained in Hermitian K-theory
plays an essential role in the holomorphic description of D-branes
for orientifold. (This is essentially a part of the
proposal in \cite{talk}.)

Perhaps one of the most interesting problems for future work
is the connection to large volume.
As discussed in the introduction, for oriented strings, 
Landau-Ginzburg orbifolds and Calabi-Yau sigma models are
connected over the moduli space of
the complexified K\"ahler class,
and the category of matrix factorizations is equivalent 
to the derived category of coherent sheaves on the underlying algebraic 
variety. 
Given the recent understanding of the equivalence
\cite{HHP} (based on the conjecture of
\cite{johannes} and the construction of \cite{orlov}),
it is very interesting  to see such a relation in the
orientifold models, and to understand peculiar properties of
D-branes which were discovered sporadically in examples,
such as the type change via navigation through non-geometric regimes
\cite{BHHW}.
Note that the K\"ahler moduli is
projected to ``real'' locus and also we expect the (discrete)
B-field to play important roles in determining the structure of
Chan-Paton factors \cite{BH2}. (See \cite{oconifold}
for the study of realized K\"ahler moduli
from a different perspective.)

\begin{acknowledgments}

We thank Ilka Brunner, Ragnar Buchweitz, Mike Douglas, Simeon Hellerman,
Manfred Herbst, Paul Horja, Kazuo Hosomichi, Misha Kapranov, Max Karoubi, 
Maxim Kontsevich, Greg Moore, David Page, and Raul Rabadan, for discussions 
and encouragement. K.H. especially thanks Dongfeng Gao and Ezra Getzler
for collaborations in related subjects.
We wish to thank the Fields Institute for support and a fruitful
working atmosphere during the Thematic Program on The Geometry of 
String Theory, during which a major part of this work was done in the
Spring of 2005. We are also grateful to the Kavli Institute for
Theoretical Physics for hospitality during the Program on
Mathematical Structure in String Theory, where some of this
work was presented. K.H.\ would also like to thank 
Saclay and the Erwin Schr\"odinger Institute, where the paper was 
finalized. The reseach of K.H. was also supported by NSERC and Alfred 
P. Sloan Foundation. The work of J.W.\ was supported in part by the 
DOE under grant number DE-FG02-90ER40542.
\end{acknowledgments}

\end{document}